\documentclass[10pt,preprint]{aastex}

\def\aas{A\&AS}   

\begin{document}

\title{
A statistical study of multiply-imaged systems in the lensing cluster Abell 68.\altaffilmark{1,2}
}

\author{
Johan Richard\altaffilmark{3},
Jean-Paul Kneib\altaffilmark{3,4},
Eric Jullo\altaffilmark{5,4},
Giovanni Covone\altaffilmark{6},
Marceau Limousin\altaffilmark{7},
Richard Ellis\altaffilmark{3},
Daniel Stark\altaffilmark{3},
Kevin Bundy\altaffilmark{3},
Oliver Czoske\altaffilmark{8},
Harald Ebeling\altaffilmark{9}
and Genevi\`eve Soucail\altaffilmark{10}
}

\altaffiltext{1}{Data presented herein were obtained at the W.M. Keck
Observatory, which is operated as a scientific partnership among the
California Institute of Technology, the University of California and
the National Aeronautics and Space Administration. The Observatory was
made possible by the generous financial support of the W.M. Keck
Foundation.}
\altaffiltext{2}{Also based on observations collected at the Very Large Telescope 
(Antu/UT1 and Melipal/UT3), European Southern Observatory, Paranal, Chile (ESO Programs 070.A-0643, 073.A-0774), the NASA/ESA 
\textit{Hubble Space Telescope} (Program \#8249) obtained at the Space Telescope Science Institute, 
which is operated by AURA under NASA contract NAS5-26555, and the Canada-France-Hawaii Telescope.}
\altaffiltext{3}{Department of Astronomy, California Institute of
  Technology, 105-24, Pasadena, CA91125; (johan,kneib,dps,rse,kbundy)@astro.caltech.edu}
\altaffiltext{4}{OAMP, Laboratoire d'Astrophysique de Marseille
UMR 6110 Traverse du Siphon 13012 Marseille, France; Jean-Paul.Kneib@oamp.fr}
\altaffiltext{5}{European Southern Observatory, Alonso de Cordova 3107, Vitacura, Chile; ejullo@eso.org }
\altaffiltext{6}{INAF - Osservatorio Astronomico di Capodimonte, Salita Moiariello, 16, 80131 Napoli, Italy; giovanni.covone@na.astro.it}
\altaffiltext{7}{Dark Cosmology Centre - Niels Bohr Institute, University of Copenhagen, Juliane Maries Vej 30,DK-2100 Copenhagen; marceau@dark-cosmology.dk }
\altaffiltext{8}{Argelander-Institut f\"ur Astronomie
            (Founded by merging of the Institut f\"ur Astrophysik
            und Extraterrestrische Forschung, the Sternwarte, and the
            Radioastronomisches Institut der Universit\"at Bonn),
Universit\"at Bonn, Auf dem H\"ugel 71, 53121 Bonn, Germany; oczoske@astro.uni-bonn.de}
\altaffiltext{9}{Institute of Astronomy, University of Hawaii,
2680 Woodlawn Drive, Honolulu, HI 96822; ebeling@ifa.hawaii.edu}
\altaffiltext{10}{Observatoire Midi-Pyr\'en\'ees, UMR5572,
     14 Avenue Edouard Belin, 31000 Toulouse, France; soucail@ast.obs-mip.fr}

\begin{abstract}

We have carried out an extensive spectroscopic survey with the Keck and VLT 
telescopes, targeting lensed galaxies in the background of the massive cluster 
Abell 68. Spectroscopic measurements are obtained for 26 lensed images, including 
a distant galaxy at $z=5.4$. Redshifts have been determined for 5 out of 7 multiply-image 
systems. Through a careful modeling of the mass distribution in the strongly-lensed regime, 
we derive a mass estimate of 5.3$\times10^{14}\,M_{\odot}$ within 500 kpc. Our mass model
is then used to constrain the redshift distribution of the remaining multiply-imaged and 
singly-imaged sources. This enables us to examine the physical properties for a subsample 
of 7 Lyman-$\alpha$ emitters at $1.7\lesssim z \lesssim 5.5$, whose unlensed luminosities 
of $\simeq10^{41}\,\mathrm{ergs\,s}^{-1}$  are fainter than similar objects found in blank fields. 
Of particular interest is an extended Lyman-$\alpha$ emission region surrounding a highly 
magnified source at $z=2.6$, detected in VIMOS Integral Field Spectroscopy data. The 
physical scale of the most distant lensed source at $z=5.4$ is very small ($<300$ pc), similar 
to the lensed $z\sim 5.6$ emitter reported by Ellis et al. (2001) in Abell 2218. New photometric 
data available for Abell 2218 allow for a direct comparison between these two unique objects.
Our survey illustrates the practicality of using lensing clusters to probe the faint end of the 
$z\sim2-5$ Lyman-$\alpha$ luminosity function in a manner that is complementary to 
blank field narrow-band surveys. 
\end{abstract}


\section{Introduction}

The central regions of massive galaxy clusters act as powerful \textit{gravitational telescopes}, magnifying 
the light from background galaxies via the effect of strong lensing. Such magnifications can attain typical
values of 1 to 3 magnitudes in concentrated cluster cores, enabling the detection of intrinsically 
fainter sources than in unlensed field surveys. The detailed study of low luminosity galaxies at $z>2$, 
where the major fraction of star-formation activity is thought to occur, is an interesting but 
poorly-understood topic. Such galaxies can either be found through their Lyman-$\alpha$ 
emission \citep[e.g.,][]{Franx,Santos}, or through their ultraviolet continuum fluxes via the 
Lyman break techniques \citep{Kneib04,Richard}.

A prerequisite for strong lensing studies of intrinsically faint galaxies at high redshift 
is an accurate measurement of the projected mass distribution in the lens \citep{Kneib03,Gavazzi,Sand05}. Such mass models are primarily limited by the number of available multiply-imaged sources of 
known redshift. Only a few well-studied clusters like Abell 1689 \citep{Broadhurst,Halkola,Limousin07}, 
with more  than 30 multiply-imaged systems, or Abell 2218 \citep{Ebbels,Kneib96} have sufficient 
constraints to permit precise modelling of each individual dark matter clump.

Spectroscopic searches for Lyman-$\alpha$ emitters (LAEs) at high redshift usually have a better 
line flux sensitivity and span a larger redshift range  ($\Delta z\sim 4$) than those of wide-field 
narrow-band  surveys. This gain in sensitivity is even larger in strong lensing applications. 
Lensed spectroscopic surveys may also be sensitive to sources with emission lines with an 
equivalent width $W<20$\ \AA\ , smaller than those in narrow-band surveys 
\citep[e.g.,][]{Fynbo,Shimasaku}. An additional complication 
in narrow-band surveys is how interlopers are treated; confirmatory spectroscopy is usually 
necessary. By contrast, in lensed surveys, the geometrical configuration of multiply-imaged 
systems can reliably distinguish between high redshift objects and low redshift interlopers 
 \citep[see e.g.,][]{Ellis}.

As our surveys expand, a variety of types of emission line galaxies are being discovered. Of
particular interest are the extended Lyman-$\alpha$ emission sources which have been mainly 
discovered in regions of significant overdensity through deep narrow-band imaging \citep{Steidel00,Francis}.
\citet{Matsuda} have identified a large number of such giant Lyman-$\alpha$ \textit{blobs} (with a 
typical size $>50$ kpc) in a 34'$\times$ 27' field of view, demonstrating the existence of a 
continuous distribution. The origin of the extended Lyman-$\alpha$ emission in such radio-quiet 
sources may be explained by gas inflow during the early stages of galaxy formation: large amounts 
of hydrogen collapsing into the dark matter potential well will cool through Lyman-$\alpha$ radiation. 
Giant Lyman-$\alpha$ blobs may thus be the progenitors of very massive galaxies in the local 
Universe. A key issue is whether the same process is seen to occur in lower-mass objects.
A route to addressing this question is to examine the nature of smaller extended Lyman-$\alpha$ 
sources, either by long-slit or Integral Field Spectroscopy (IFS). This identification is more easily 
accomplished in strongly-lensed sources where magnification will stretch the observed physical scales.

The spatial magnification associated with lensing can also be used to yield physical sizes
for the most distant sources. Using strong lensing in the cluster Abell 2218, \citet[]{Ellis} located
a remarkably small source at $z$=5.6 where the combination of the Ly$\alpha$ emission line
flux density and the weak stellar continuum were used to deduce a young age and modest stellar
mass ($\simeq\ 10^{6-7}\,M_{\odot}$) consistent, perhaps, with a forming globular cluster. 
Further surveys are required to evaluate whether such systems are common at $z\simeq$6.


The major drawback arising from the study of lensed sources located through studies of individual 
clusters is, of course, the significant cosmic variance that is associated with the small volumes
being probed. Compared to field surveys, any statistical inferences on the abundances of various
classes of populations may be much more uncertain, even granting fainter sources are probed.
To overcome this limitation, an effective survey would have to be conducted through a 
large sample ($\simeq$20-40) of lensing clusters, each with reliable mass models based on the
spectroscopic study of many multiply-imaged systems \citep{Kneib96}. Fortunately, the construction
of such a sample of well-mapped clusters is now a realistic proposition. Several Hubble
Space Telescope (HST) snapshot imaging surveys of X-ray luminous clusters are now underway 
with associated ground-based spectroscopy, such as the MAssive Clusters Survey 
(MACS, GO\#10491, P.I.: H. Ebeling) and the  Local Cluster Substructure Survey 
(LoCuSS, GO\#10881, P.I.: G. Smith). 

The purpose of this paper is to illustrate the promise of such surveys by examining spectroscopically
the rich population of lensed sources located in the lensing cluster Abell 68  ($\alpha$=00:37:06.81 
$\delta$=+09:09:24.0 J2000, $z=0.255$), one of the most X-ray luminous clusters 
($L_{\mathrm{X}}\sim 8.4\pm 2.3\times 10^{44}\mathrm{\,erg\,s}^{-1}$, 0.1--2.4 keV) in the
X-ray Brightest Abell-type Clusters sample (XBACS, \citet{Ebeling}).
Strong lensing in this cluster has been previously studied by \citet{Smith05}, hereafter S05, 
as part of a survey of 10 X-ray luminous galaxy clusters at $z\sim 0.2$. Smith et al identified a list of 
potential multiple-image systems, a few of which were confirmed spectroscopically. Here we
significantly extend this work by securing the redshifts of new multiple-image systems, 
many of which are strongly-lensed Lyman-$\alpha$ emitters at $z\gtrsim 2$. The combination 
of a large magnification factor, high-resolution HST imaging and broad-band photometry 
enables us to demonstrate the value of studying the physical properties of these faint emitters, 
such as their star formation rates, intrinsic scales and stellar masses. The paper is intended 
to illustrate the significant promise of continuing such spectroscopic work with the larger samples 
of clusters now being surveyed with HST.

The paper is organized as follows. In Section 2, we describe the various observations and the 
reduction of the spectroscopic data. We present in Section 3 the strong-lensing constraints, in the light 
of the redshifts and identification of new multiply-imaged systems. Section 4 presents
a mass model of the cluster from which the source magnifications are deduced. The physical properties 
of the various categories of high redshift LAEs are presented in Section 5 and the implications are
discussed in the context of the limitations of blank field surveys in Section 6. We summarize our conclusions 
in Section 7.

Throughout this paper, we adopt the following cosmology: a flat $\Lambda$-dominated
Universe with the values $\Omega_{\Lambda}=0.7$, $\Omega_{m}=0.3$,
$\Omega_{b}=0.045$ and $H_{0}=70\ \mathrm{km\,s}^{-1}\,\mathrm{Mpc}^{-1}$. All
magnitudes given in the paper are quoted in the $AB$ system \citep{AB}. The correction values $
C_\mathrm{AB}$ between $\mathrm{AB}$ and Vega photometric systems, defined as 
$m_\mathrm{AB}=m_{\mathrm{Vega}}+C_\mathrm{AB}$, are reported in Table \ref{imaging} for each filter. 
At the redshift $z=0.255$ of the cluster, the angular diameter distance is 3.9 kpc\ arcsec$^{-1}$.

\section{Observations and Data Reduction}

\begin{figure*}
\centerline{\mbox{\plotone{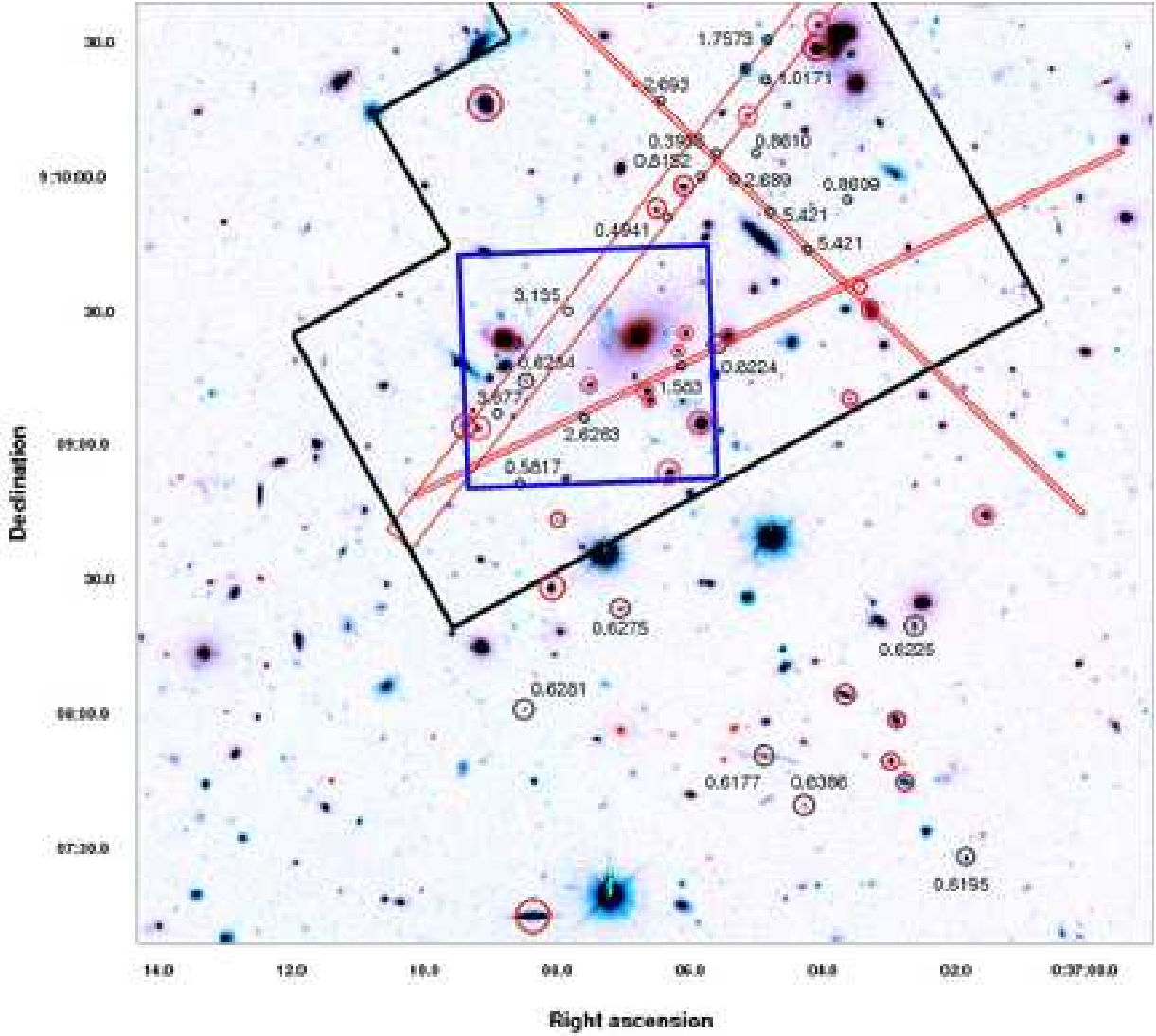}}}
\caption{\label{field}Composite CFH12k-$BRI$ color image of the field of view around the 
center of Abell 68. We overplot the redshift measurements obtained for galaxies located 
in the background of the cluster (black labels). Red circles represent cluster members 
confirmed with spectroscopy. We delineate the imprints of the HST/WFPC2 (black polygon) 
and the VIMOS/IFU (blue square) fields, as well as the spatial coverage of the different 
LRIS long-slit configurations (red rectangles).}
\end{figure*}

We present in this section the photometric and spectroscopic datasets used to assemble our 
catalog. High resolution images are crucial for the morphological identification of multiple image 
systems and the precise astrometric position of the sources studied here, whereas multicolor images 
are used to estimate their spectral energy distributions. Redshift and emission lines measurements 
for individual objects were obtained during subsequent spectroscopic observations. These included 
multi-object spectroscopy of multiply-imaged candidates, as well as systematic long-slit searches 
in the central regions of the cluster. Figure \ref{field} shows the location of the main spectroscopic 
settings in the cluster field.

\subsection{Imaging data}
\label{images}

A considerable body of multi-wavelength data exists in the field around Abell 68, including high resolution 
HST imaging. The main characteristics of the dataset used in this study are summarized in Table \ref{imaging}. 3 $\times$ 2.5 ks of integration time with the Wide Field Planetary Camera (WFPC2) was 
obtained during Cycle 8 in the $R$-F702W band, as part of HST program \#8249 (PI : J.P. Kneib). 
Observations were carried out in low sky mode, and a 1.0\arcsec\ dithering pattern was used between 
each exposure. Details on the reduction of these data are given in S05.\\

Recognition of faint multiply-imaged systems in the vicinity of the cluster core is hindered by the
dominant stellar halo of the Brightest Cluster Galaxy (BCG). To overcome this, we fitted and 
subtracted from the HST image a model representation of the surface brightness distribution using 
the IRAF task {\tt ellipse}. Both the position angle and ellipticity were allowed to vary as a function of 
the semimajor axis in the fitted elliptical isophotes, as well as the isophote centroid in the central 
part. This procedure was found to give satisfactory residuals at the center (Figure \ref{multiple}).\\

Associated optical images in $B,R,I$  have been obtained on UT 1999 November 19 
using the CFH12k camera at CFHT. These sample a field of 42'$\times$28' at a 0.205\arcsec\ 
pixel scale. The total exposure times are 8.1, 7.2 and 3.6 ks in the $B$, $R$ and $I$ band, 
respectively. The data was  reduced using procedures similar to those described by 
\citet{Czoske02} and \citet{Bardeau}.

At longer wavelengths, Abell 68 has been observed at the Very Large Telescope using the 
FOcal Reducer / low dispersion Spectrograph (FORS2/UT4) in the $z$-band on UT 2002 
October 06, and the Infrared Spectrometer And Array Camera (ISAAC/UT1) in the $J$ 
and $H$ bands on UT 2002 September 29 . The field of view of the FORS2 image is 
7.2 $\times$ 7.2 arcmins after dithering, with a pixel size of 0.252\arcsec, and we used 
80 dithered exposures of 120 s. The field of view of the ISAAC images is about 2.5 
$\times$ 2.5 arcmins after dithering, with a pixel size of 0.148\arcsec, the subintegration 
$\times$ integration times of the dithered exposures were 6 $\times$ 35 s and 
10 $\times$ 12 s in the $J$ and $H$ bands, respectively. All these data have been 
reduced using procedures similar to those described by \citet{Richard}.

\begin{table}
\begin{tabular}{lcclccc}
\tableline
\tableline
Instrument&Filter & Exposure Time & Pixel Size & Depth & $C_\mathrm{AB}$ & Seeing\\
& & (ks) & \arcsec & $AB$ mag. & mag. & \arcsec \\
\tableline
CFH12k & $B$ & 8.1  & 0.206   & 27.4 & -0.066 & 1.11 \\
CFH12k & $R$ & 7.2  & 0.206   & 27.2 & 0.246  & 0.67\\
WFPC2 & $R_{702W}$ & 7.5 & 0.1 & 28.0 & 0.299  & 0.17 \\
CFH12k & $I$ & 3.6  & 0.206   & 26.5 & 0.462  & 0.58\\
FORS2  & $z$ & 9.6  & 0.252   & 26.5 & 0.554  & 0.71\\
ISAAC  & $J$ & 6.48 & 0.148   & 26.2 & 0.945  & 0.48\\
ISAAC  & $H$ & 7.12 & 0.148   & 26.3 & 1.412  & 0.48\\
\tableline
\end{tabular}
\caption{\label{imaging}Properties of the photometric dataset: from left to 
right: instrument and filter names, total integration time, pixel size, photometric depth 
(defined as 4 pixels above 3$\,\sigma$, where $\sigma$ stands for the typical 
local background noise), photometric correction $C_\mathrm{AB}$ between $AB$ 
and Vega systems, seeing measured on bright unsaturated stars.}
\end{table}

\subsection{Keck Multi-Object Spectroscopy}

The Low Resolution Imaging Spectrograph (LRIS; \citet{lris}) on Keck I has been used in 
multi-object (MOS) mode during two observing runs, in order to target background galaxies 
and multiply-imaged candidates selected on the basis of  morphology and colors.

On UT 2001 August 4, 4 exposures of 1.8 ks were acquired with a 31 slits-mask, using a 
300 l\ mm$^{-1}$ grism blazed at 5000 \AA\ in the single red channel of the camera, which 
covers the approximate range 5500-9900 \AA\ at the dispersion of 2.5 \AA\ per pixel. The 
average seeing was $\sim 1.0$\arcsec. The night was photometric and spectrophotometric 
standard stars were used for flux calibration.

On UT 2002 November 30, a 32 slits-mask was used during 3 $\times$ 2.4 ks of integration 
time. A 6800 \AA\ dichroic separated the red channel of the instrument, equipped with a 600 l 
mm$^{-1}$ grating blazed at 7500 \AA , from the blue channel equipped with a 400 l\ mm$^{-1}$ 
blazed at 3400 \AA . The whole setting covers the wavelength range 3500-9500 \AA\ with 
dispersions of 1.28 and 1.09 \AA\ per pixel in the red and blue channels, respectively. Despite 
good seeing conditions ($\sim 0.8$\arcsec), the night was not photometric and no standard stars 
were observed.

These datasets were reduced using standard IRAF procedures for bias removal, flat-fielding, 
wavelength and flux calibration, sky subtraction and extraction of the one-dimensional spectra. 

\subsection{Keck long-slit spectroscopy}

\subsubsection{Optical spectroscopy}

Abell 68 was observed on UT 2002 September 11 with LRIS, in the course of a survey 
targeting low-luminosity Lyman-$\alpha$ sources at high redshift  \citep{Santos}. A 
175\arcsec-long and 1\arcsec-wide slit was used to map the high magnification regions 
of a sample of $z\sim 0.2$ lensing clusters. In the case of Abell 68, 6 adjacent slit settings 
scanned the theoretical location of the critical lines at $z\sim 5$, with 2$\times$1000 s of 
integration time at each  position. The reduction of these data is detailed in \citep{Santos}.

In addition to the detection of high redshift sources through their Lyman-$\alpha$ 
emission, this \textit{blind} spectroscopic survey provided secure redshifts for a number of 
lensed background galaxies serendipitously falling into the long slit. 

On UT 2003 August 26, a single long-slit LRIS position was aligned on two components 
of a triply-imaged system, discovered as R-dropouts by a $RIz$ color-color selection technique 
(see Sect. \ref{systems}). A 300 l\ mm$^{-1}$ grism blazed at 5000 \AA\ and a 
600 l\ mm$^{-1}$ grating blazed at 1$\mu$m were used in the blue and red channels of the 
instrument, both lightpaths being separated by a dichroic at 6800 \AA. Two exposures of 1.2 ks 
were acquired at this position, with a  5\arcsec\ dithering offset along the slit. 

Finally, an additional LRIS long-slit integration of 3 $\times$ 1.2 ks was acquired on UT 2005 
November 29 with a 600 l\ mm$^{-1}$ grism blazed at 4000 \AA, a 400 l\ mm$^{-1}$ grating 
blazed at 8500 \AA, a 5600 \AA\ dichroic, and 5\arcsec\ dithering offsets.

\subsubsection{Near Infrared Spectroscopy}

Abell 68 was observed on UT 2005 October 13 using the Near InfraRed SPECtrograph 
(NIRSPEC, \citet{nirspec}) on Keck II, during a spectroscopic survey of the critical lines
of lensing clusters similar to the LRIS survey described before, but at longer
wavelengths \citep{Stark}. A 42\arcsec $\times$ 0.76\arcsec\ long slit was used at 2 adjacent
slit positions, with 9 $\times$ 600 s of integration time on each of them. The spectra were reduced 
using IDL routines, following optimal spectroscopic reductions techniques presented in \citet{Kelson}. 
More details are presented in a forthcoming paper (Stark et al.\ 2006b, in press).

\subsection{VLT - Integral Field Spectroscopy}

Abell 68 was observed on UT 2004 August 12 using the VIsible Multi-Object Spectrograph 
(VIMOS, \citet{Lefevre03}) on VLT/UT3 in low-resolution (LR-Blue grism) Intregral Field 
Spectroscopy mode, as part of a survey targeting the central regions of an intermediate-redshift 
galaxy cluster sample (073.A-0774, PI: Soucail). The 54\arcsec $\times$ 54\arcsec\ field of view of the Integral Field Unit (IFU, composed of 6400 fibers, splitted in 4 quadrants feeding the 4 VIMOS CCDs) was centered on the cD galaxy. In the given configuration, the spectral resolution is about $200$ and the 
diameter of the fibers is $0.66$\arcsec, covering the wavelength range 3900-6800 \AA\ 
with a dispersion of 5.355 \AA\ per pixel. 2 $\times$ 2.4 ks of integration time have been acquired 
without dithering.

These 3D spectroscopic data have been reduced using the 
Vimos Interactive Pipeline Graphical Interface \citep[VIPGI,][]{vipgi}\footnote{VIPGI has been 
developed within the VIRMOS Consortium. For more information, see 
http://cosmos.mi.iasf.cnr.it/pandora/vipgi.html}.  Before building the data cube for each 
exposure, every step in the data reduction was performed on a single quadrant basis.
After bias subtraction and cosmic ray removal (see \citet{Zanichelli} for a description of 
the algorithm), the spectra were traced on the CCDs with the help of the high S/N spectra 
of a continuum lamp. Following wavelength calibration, inhomogeneities in fiber
efficiencies were corrected by measuring the counts in the 5577 \AA\ sky emission line 
after subtracting the contribution from a galaxy spectrum where fibers cover a galaxy position.
The flux-calibration is applied by using observations of a standard star in each quadrant.
See \citet{Covone}, who present similar data on Abell 2667, for a detailed description 
of the procedures.

\subsection{Redshift measurements}

\begin{table*}
\footnotesize
\begin{tabular}{llllllllll}
\tableline
\tableline
ID & RA & Dec & $R$ & $z$ & $C$ & Features & $\mu$ & W$_0$ \\
& 00: & 09: & [mag] & & & & [mag] & [\AA] \\
\tableline
&&&&&&&& ($Ly\alpha$) \\
1 (C15a) & 37:04.297 & 09:43.40 & $26.00 \pm 0.18$ & $ 5.421 \pm 0.0021$ & 9 & $Ly\alpha$ & 2.74$\pm0.08$ & $ 53 \pm 16$\tablenotemark{a}\\
2 (C15b) & 37:04.861 & 09:51.78 & $26.52 \pm 0.22$ & $ 5.421 \pm 0.0021$ & 9 & $Ly\alpha$ & 2.89$\pm0.07$ & \\
3 (C26) & 37:08.960 & 09:06.90 & $27.24 \pm 0.25$ & $ 3.677 \pm 0.0022 $ & 9 & $Ly\alpha$ & 2.06$\pm0.03$ &  $ 107 \pm 8.2 $\\
4 (C23a) & 37:07.902 & 09:29.80 & $26.44 \pm 0.18$ & $ 3.135 \pm 0.0021 $ & 9 & $Ly\alpha$ & 2.47$\pm0.07$ & $25.9 \pm 5.2 $\\
5 (C25) & 37:06.506 & 10:16.70 & $24.40 \pm 0.07$ & $ 2.6930 \pm 0.0021$ & 9 & $Ly\alpha$ & 1.12$\pm0.07$ & $ 10.4 \pm 5.1 $\\
6 (C20c) & 37:05.405 & 09:59.14 & $25.15 \pm 0.09$ & $ 2.6890 \pm 0.0020$ & 9 & $Ly\alpha$ & 3.61$\pm0.09$ & $ 30.4 \pm 6.3 $\\
7 (C4) & 37:07.657 & 09:05.90 & $23.31 \pm 0.04$ & $ 2.6280 \pm 0.0021$ & 9 & $Ly\alpha$ & 4.15$\pm0.16$ &  $ 42.4 \pm 2.3 $\\
8 (C27) & 37:04.906 & 10:30.20 & $22.81 \pm 0.03$ & $ 1.7546 \pm 0.0021$ & 4 & $Ly\alpha$ & 1.72$\pm0.10$ & $ 2.4 \pm 0.5 $ \\
\tableline
&&&&&&&& ($[O_{II}]$) \\
9  (C1a) & 37:06.207 & 09:17.49 & $24.02 \pm 0.04$ & $ 1.5836 \pm 0.0011$ & 4 & $[O_{III}]$, H$\beta$ & 2.52$\pm 0.06$ & $ < 50 $\\
10 (C12) & 37:04.930 & 10:21.40 & $21.67 \pm 0.02$ & $ 1.0171 \pm 0.0011$ & 3 & $[O_{II}]$, H$\alpha$ & 1.63$\pm0.06$ & $ 10.5 \pm 1.4 $\\
11 (C7) & 37:05.073 & 10:04.90 & $22.78 \pm 0.03$ & $ 0.8610 \pm 0.0007$ & 3 & $[O_{II}]$ & 1.59$\pm0.02$ & $ 21.3 \pm 5.9 $\\
12 (C8) & 37:03.700 & 09:54.10 & $23.08 \pm 0.02$ & $ 0.8609 \pm 0.0008$ & 4 & $[O_{II}]$ & 1.37$\pm0.03$ & $ 112.8 \pm 5.1 $\\
13 (C24) & 37:05.900 & 09:59.70 & $23.64 \pm 0.02$ & $ 0.8152 \pm 0.0007$ & 1 & K, H, H$\gamma$ & 1.33$\pm0.01$ &  N/A \\
14 & 37:04.352 & 07:39.60 & $\it 21.91 \pm 0.03$ & $ 0.6386 \pm 0.0007$ & 3 & $[O_{II}]$ & 0.10 & $ 14.1 \pm 4.5 $\\
15 & 37:08.541 & 08:01.30 & $\it 22.49 \pm 0.03$ & $ 0.6281 \pm 0.0006$ & 3 & K, H & 0.20 & $ < 10 $\\
16 & 37:07.100 & 08:23.10 & $\it 21.44 \pm 0.02$ & $ 0.6275 \pm 0.0006$ & 3 & K, H & 0.28 & $ 14.6 \pm 2.2 $\\
17 (C14) & 37:08.534 & 09:14.10 & $22.02 \pm 0.02$ & $ 0.6234 \pm 0.0007$ & 2 & K, H & 0.93$\pm0.01$ & $ 16.5 \pm 7.4 $\\
18 & 37:02.707 & 08:19.18 & $\it 20.52 \pm 0.02$ & $ 0.6225 \pm 0.0006$ & 3 & K, H & 0.20 & $ 6.7 \pm 1.3 $\\
19 & 37:05.505 & 09:24.24 & $22.71 \pm 0.02$ & $ 0.6224 \pm 0.0006$& 4 & $[O_{II}]$, $[O_{III}]$ & 1.83 & $ 25.3 \pm 6.5 $\\
20 & 37:01.890 & 07:27.70 & $\it 21.40 \pm 0.03$ & $ 0.6195 \pm 0.0005$ & 4 & $Mg_{II}$, $[O_{III}]$ & 0.10 & N/A\\
21 & 37:05.000 & 07:50.60 & $\it 20.33 \pm 0.02$ & $ 0.6177 \pm 0.0007$ & 4 & $[O_{II}]$ & 0.10 & $ 10.7 \pm 1.7 $\\
22 & 37:01.400 & 05:55.20 & $\it 22.10 \pm 0.02$ & $ 0.5944 \pm 0.0005$ & 4 & H$\beta$, $[O_{III}]$ & 0.0 & N/A\\
23 & 37:08.620 & 08:51.40 & $23.29 \pm 0.04$ & $ 0.5817 \pm 0.0006$ & 4 & H$\beta$, $[O_{III}]$ & 0.51 &  $ < 10 $\\
24 & 37:06.410 & 09:50.65 & $24.66 \pm 0.05$ & $ 0.4941 \pm 0.0006$ & 4 & H$\beta$, $[O_{III}]$ & 0.75 & $ < 10 $\\
25 & 37:05.670 & 10:04.90 & $24.00 \pm 0.04$ & $ 0.3958 \pm 0.0007$ & 4 & H$\alpha$, $[O_{III}]$, $[O_{II}]$ & 0.51 & $ 24.5 \pm 2.2 $\\
26 & 36:57.170 & 07:00.80 & $\it 21.03 \pm 0.01$ & $ 0.3693 \pm 0.0005$ & 4 & $[O_{II}]$, $[O_{III}]$ & 0.0 &  $ 10.2 \pm 2.1 $\\
\tableline
\end{tabular}

\caption{\label{speccat}Spectroscopic catalog of lensed background galaxies. From 
left to right: astrometric position (J2000), $R$-band magnitude in $AB$ system (WFPC/F702W, 
or CFH12k (in italic)), spectroscopic redshift, redshift confidence class (see text for details), 
main spectroscopic features, magnification factor (in magnitudes), rest-frame equivalent 
width of Lyman-$\alpha$ (top half) or $[O_{II}]$ emission lines (bottom half).}

\tablenotetext{a}{Measured on the averaged spectrum of C15a and C15b}
\end{table*}

\begin{figure*}
\setlength{\parindent}{0.cm}

\begin{minipage}{2.8cm}
\centerline{\mbox{\includegraphics[width=0.95\textwidth]{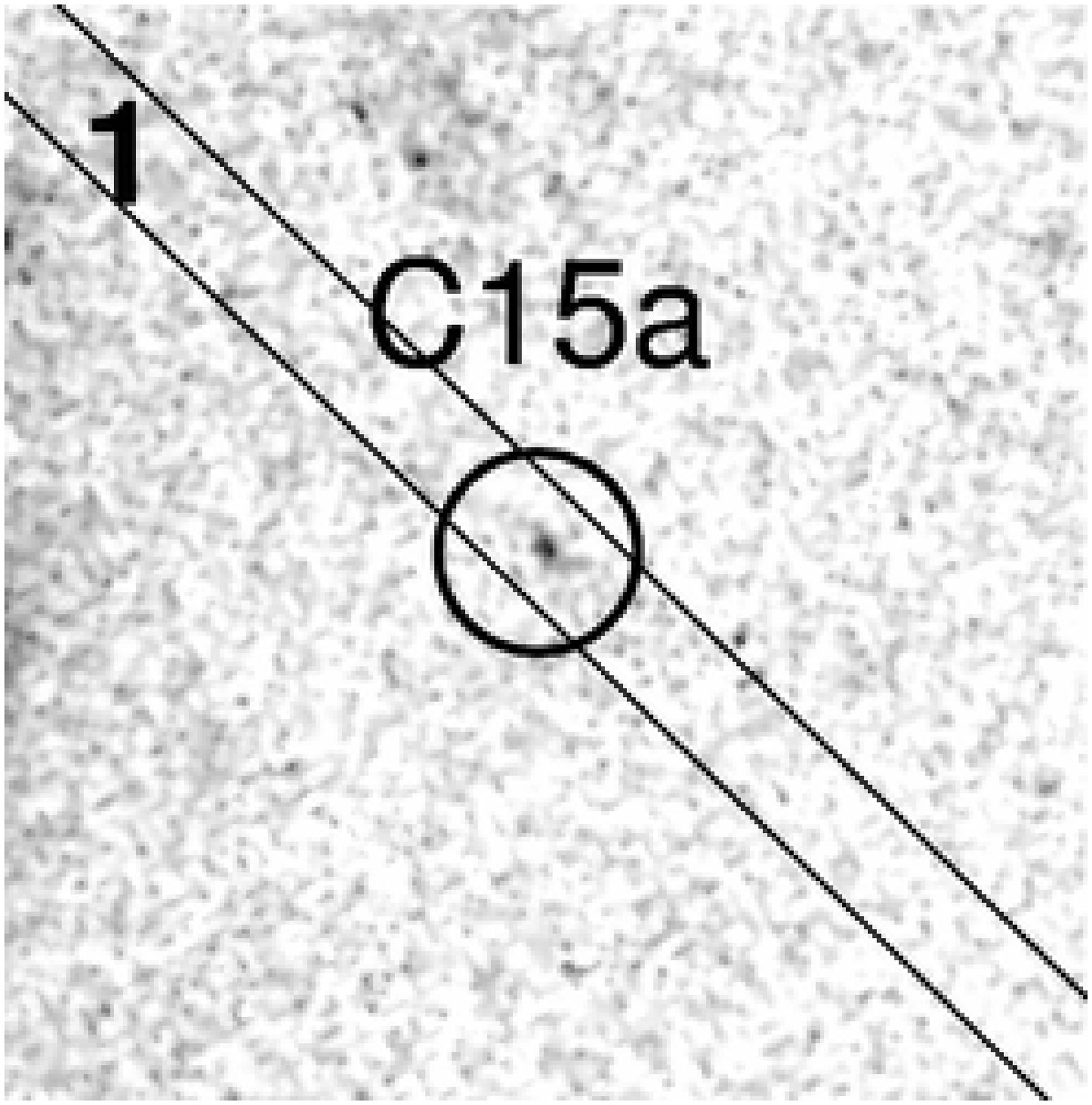}}}
\end{minipage}
\begin{minipage}{5.cm}
\centerline{\mbox{\includegraphics[height=0.95\textwidth,angle=270]{f2b.eps}}}
\end{minipage}
\begin{minipage}{2.8cm}
\centerline{\mbox{\includegraphics[width=0.95\textwidth]{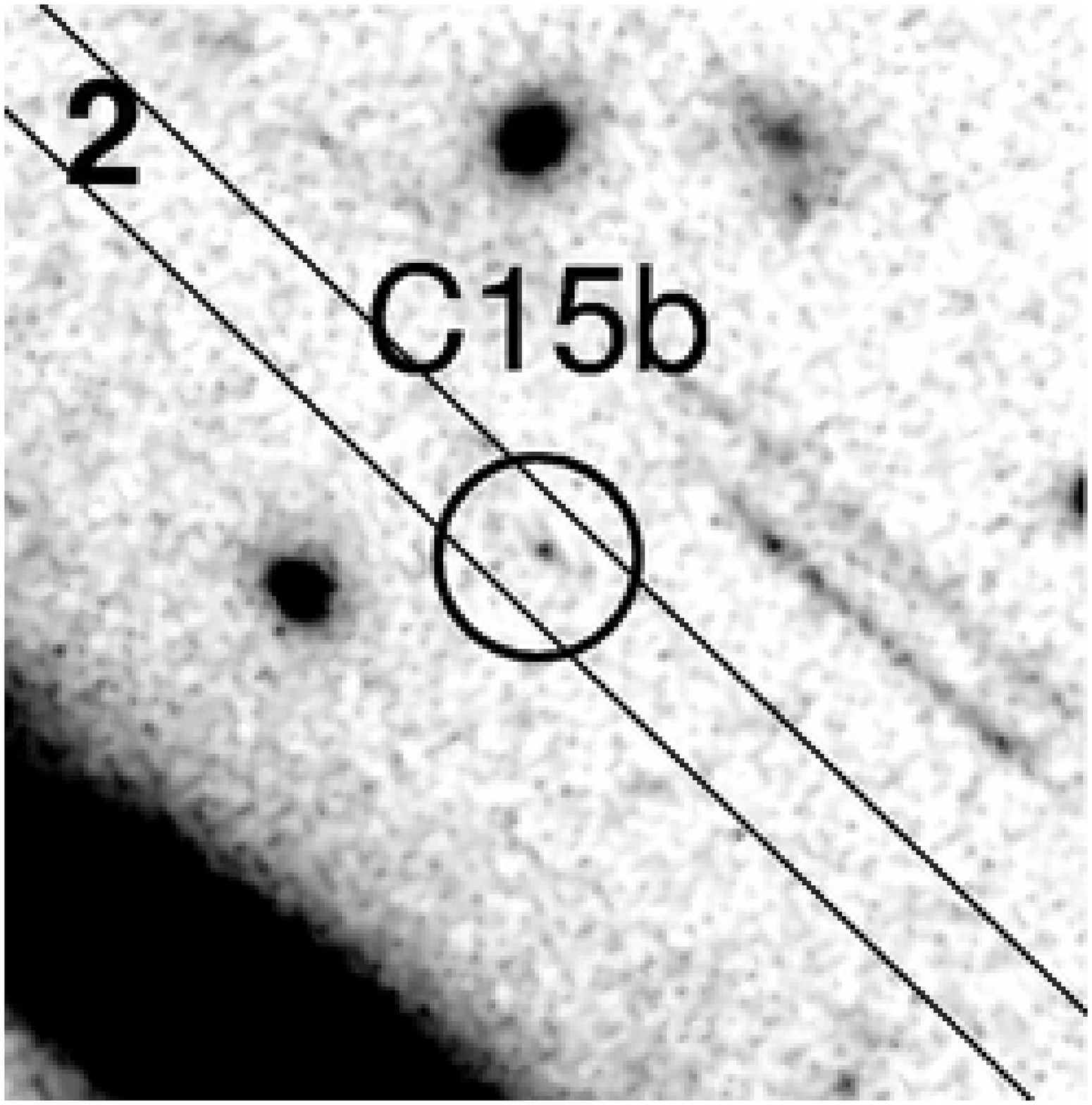}}}
\end{minipage}
\begin{minipage}{5.cm}
\centerline{\mbox{\includegraphics[height=0.95\textwidth,angle=270]{f2d.eps}}}
\end{minipage}
\\
\begin{minipage}{2.8cm}
\centerline{\mbox{\includegraphics[width=0.95\textwidth]{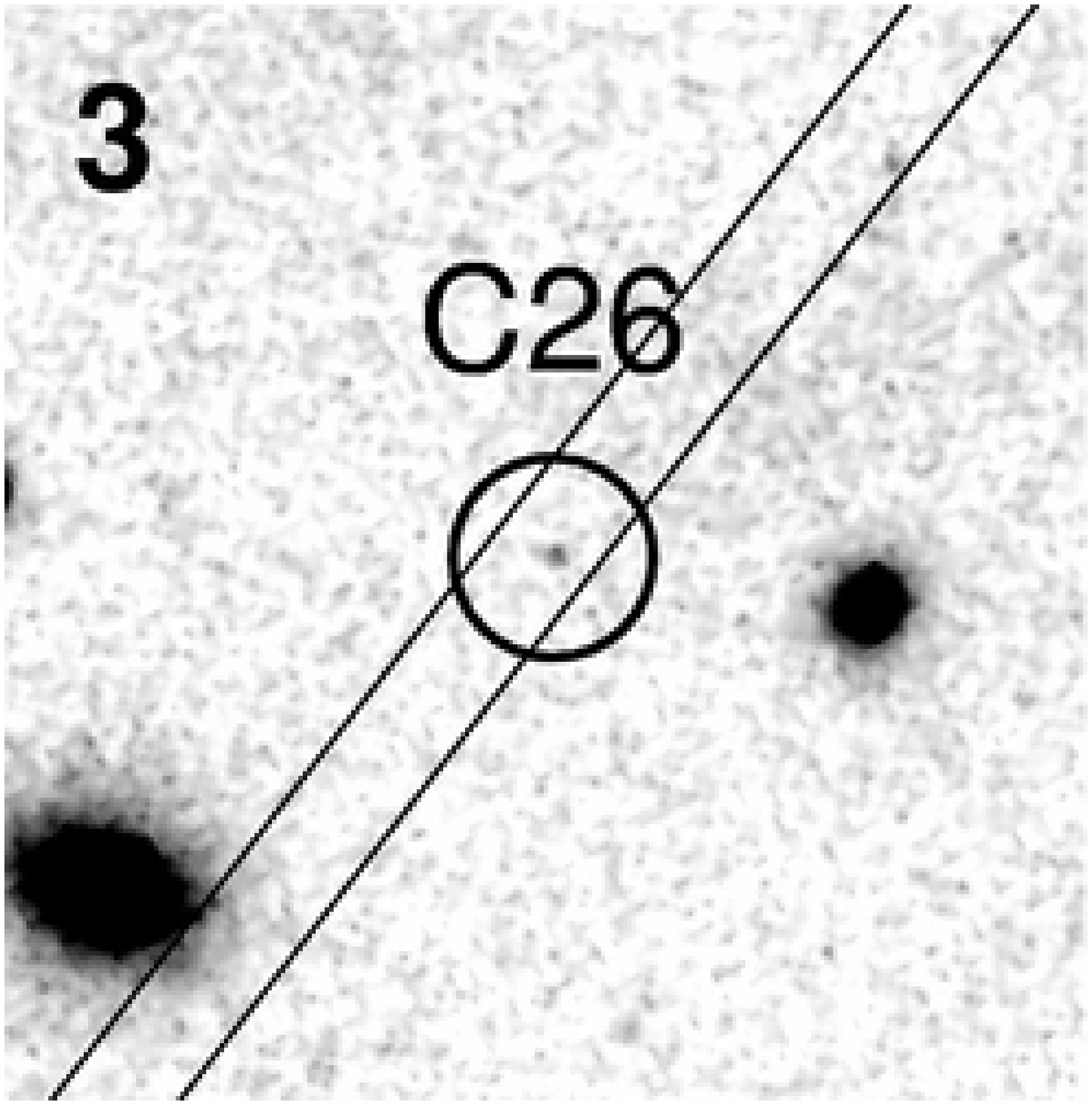}}}
\end{minipage}
\begin{minipage}{5.cm}
\centerline{\mbox{\includegraphics[height=0.95\textwidth,angle=270]{f2f.eps}}}
\end{minipage}
\begin{minipage}{2.8cm}
\centerline{\mbox{\includegraphics[width=0.95\textwidth]{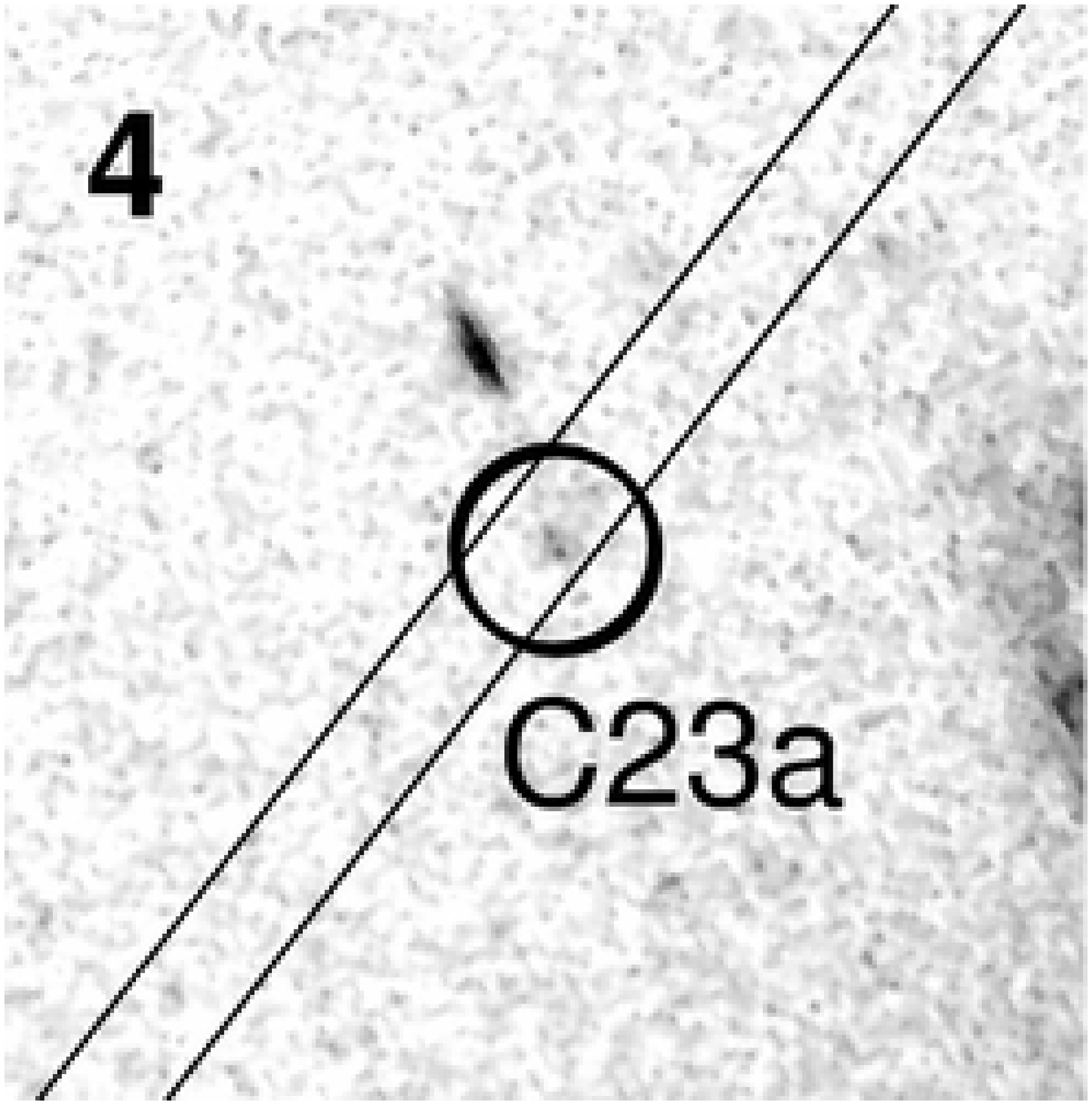}}}
\end{minipage}
\begin{minipage}{5.cm}
\centerline{\mbox{\includegraphics[height=0.95\textwidth,angle=270]{f2h.eps}}}
\end{minipage}
\\
\begin{minipage}{2.8cm}
\centerline{\mbox{\includegraphics[width=0.95\textwidth]{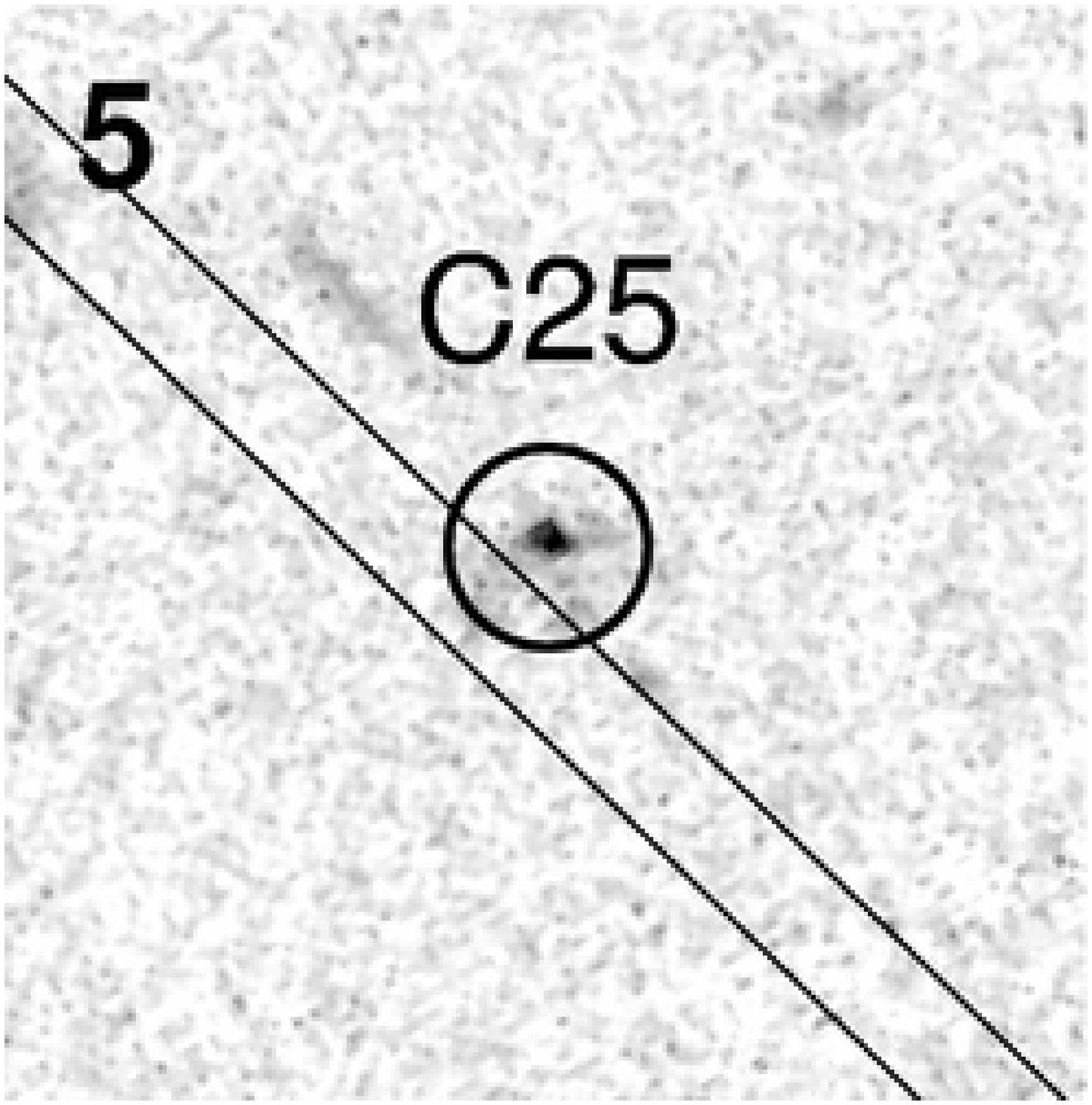}}}
\end{minipage}
\begin{minipage}{5.cm}
\centerline{\mbox{\includegraphics[height=0.95\textwidth,angle=270]{f2j.eps}}}
\end{minipage}
\begin{minipage}{2.8cm}
\centerline{\mbox{\includegraphics[width=0.95\textwidth]{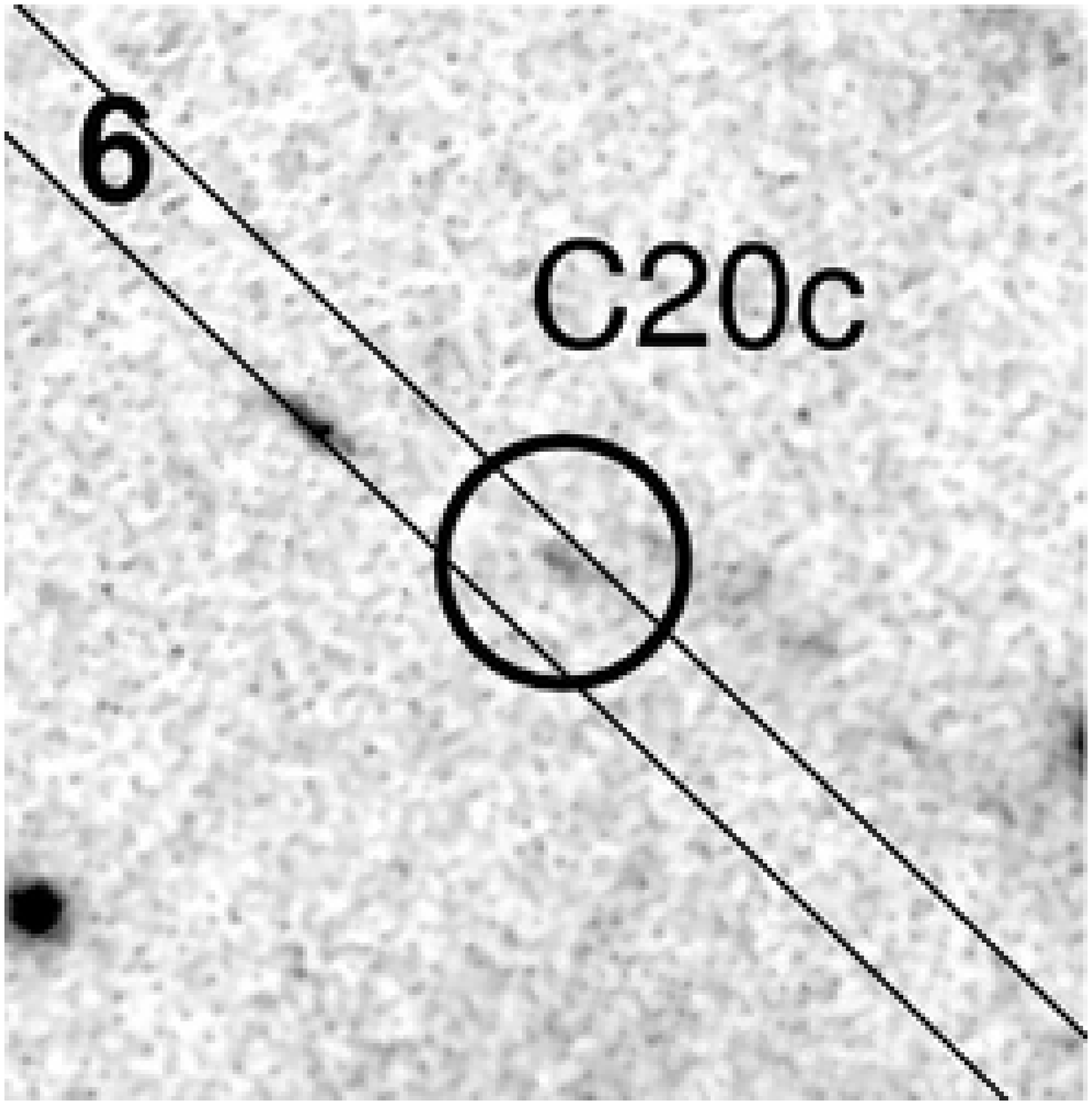}}}
\end{minipage}
\begin{minipage}{5.cm}
\centerline{\mbox{\includegraphics[height=0.95\textwidth,angle=270]{f2l.eps}}}
\end{minipage}
\\
\begin{minipage}{2.8cm}
\centerline{\mbox{\includegraphics[width=0.95\textwidth]{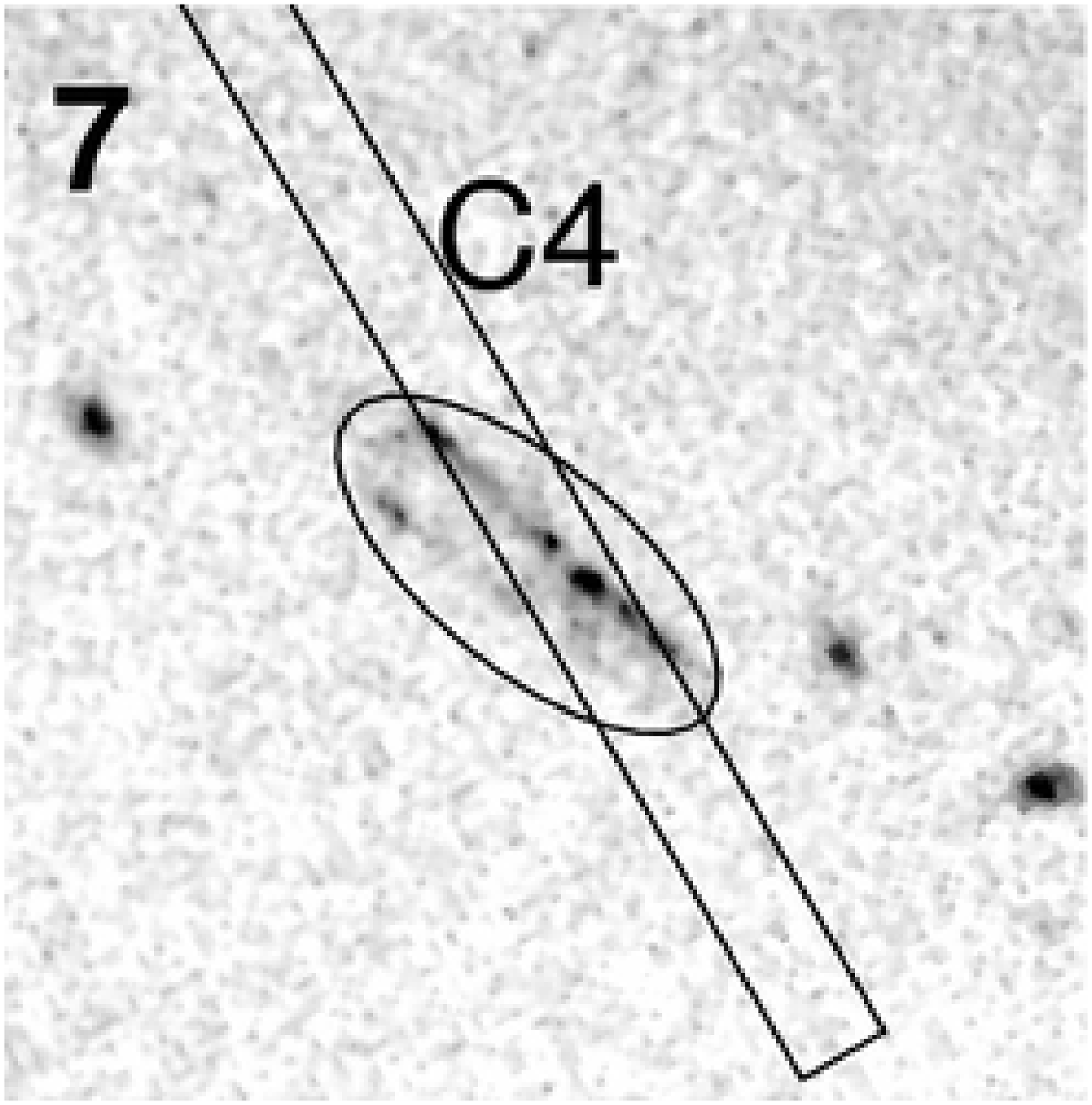}}}
\end{minipage}
\begin{minipage}{5.cm}
\centerline{\mbox{\includegraphics[height=0.95\textwidth,angle=270]{f2n.eps}}}
\end{minipage}
\begin{minipage}{2.8cm}
\centerline{\mbox{\includegraphics[width=0.95\textwidth]{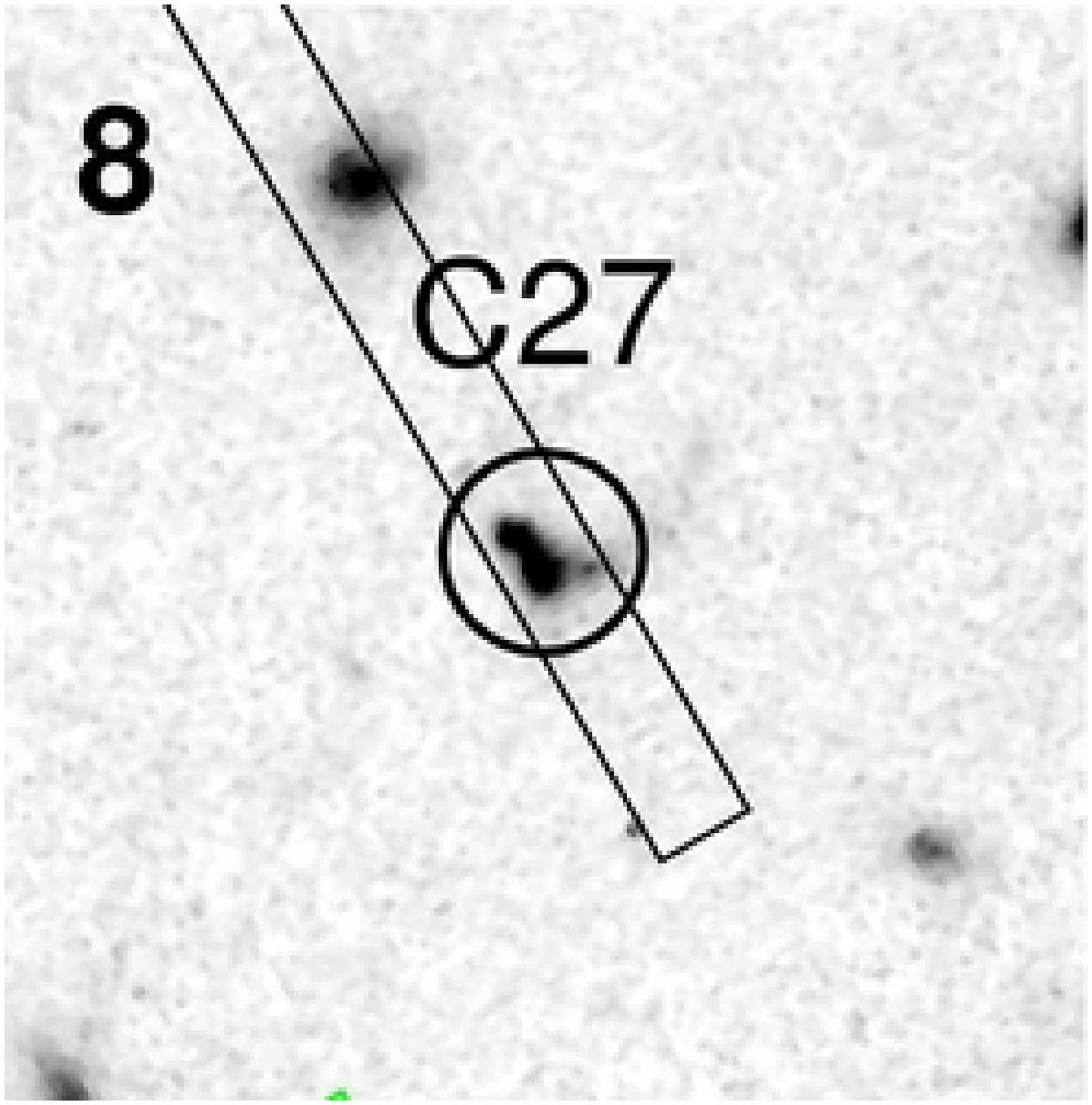}}}
\end{minipage}
\begin{minipage}{5.cm}
\centerline{\mbox{\includegraphics[height=0.95\textwidth,angle=270]{f2p.eps}}}
\end{minipage}
\\
\begin{minipage}{2.8cm}
\centerline{\mbox{\includegraphics[width=0.95\textwidth]{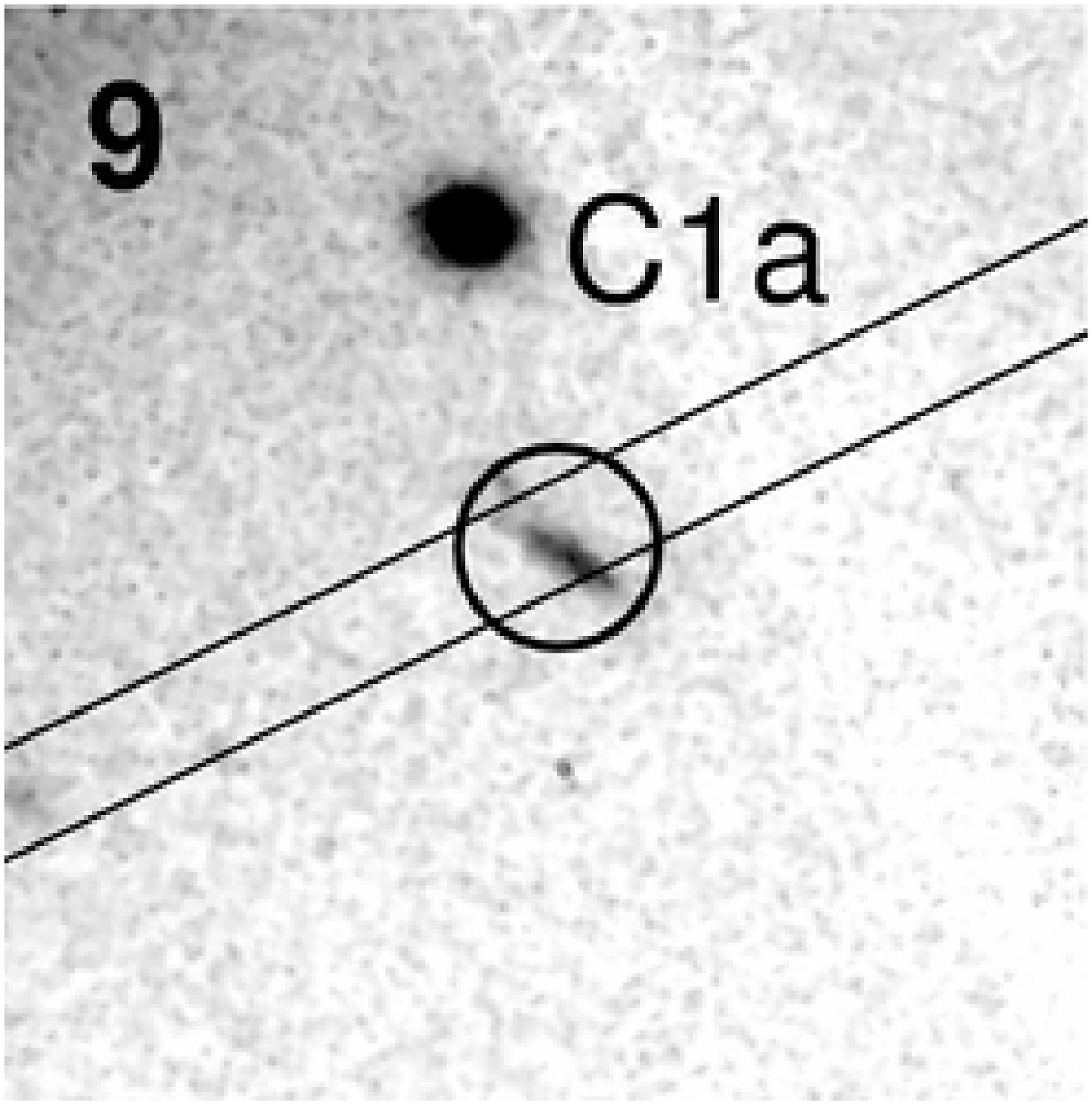}}}
\end{minipage}
\begin{minipage}{5.cm}
\centerline{\mbox{\includegraphics[height=0.95\textwidth,angle=270]{f2r.eps}}}
\end{minipage}
\begin{minipage}{2.8cm}
\centerline{\mbox{\includegraphics[width=0.95\textwidth]{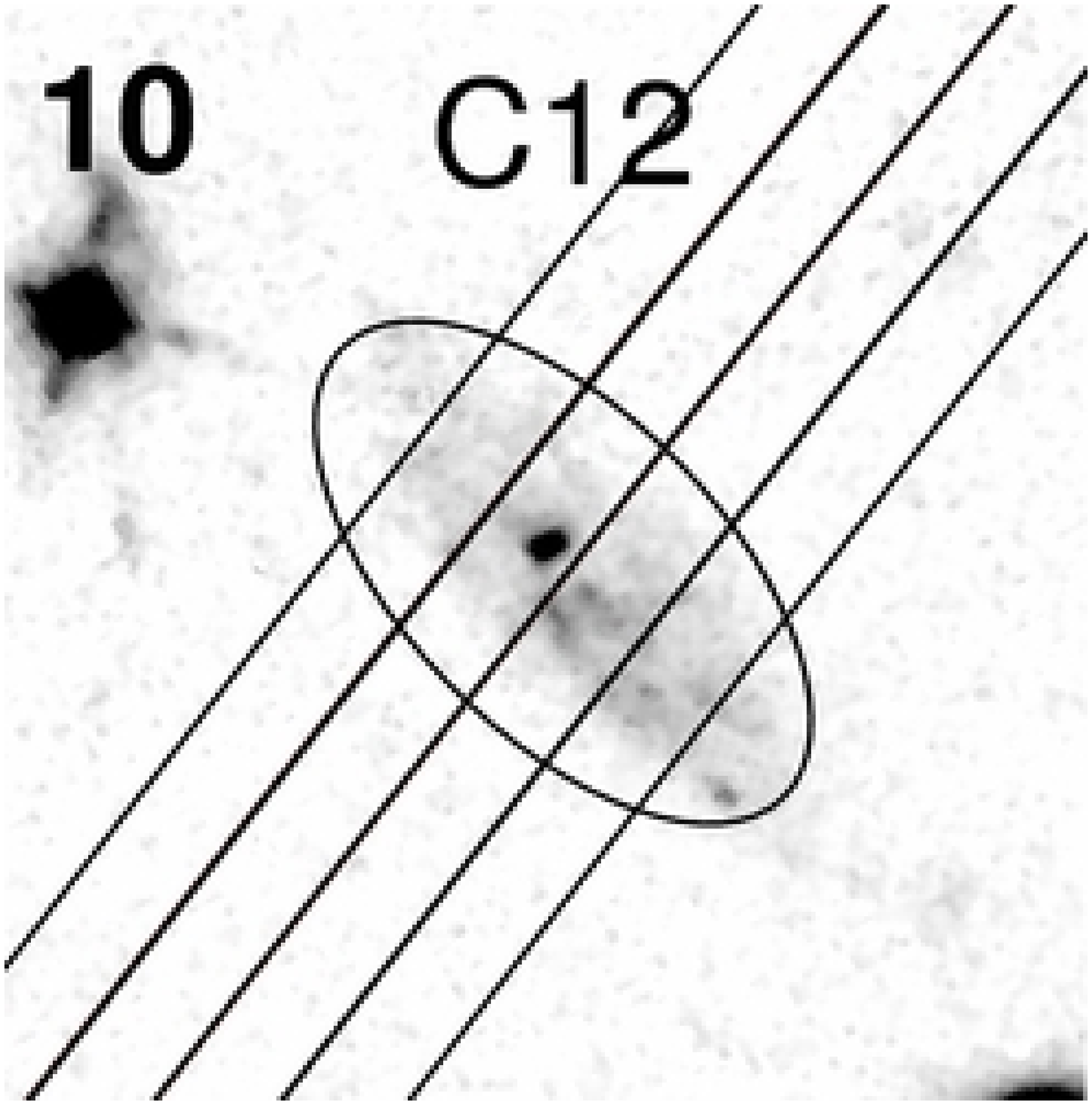}}}
\end{minipage}
\begin{minipage}{5.cm}
\centerline{\mbox{\includegraphics[height=0.95\textwidth,angle=270]{f2t.eps}}}
\end{minipage}
\\
\caption{\label{spec1d}10\arcsec $\times$ 10\arcsec\ WFPC/F702W zoomed images with the 
slit location and extracted spectra for all individual $z>1$ sources. Dotted lines outline the 
emission and absorption features identified in each spectrum. Upper red spectra have 
been smoothed using a $\sigma=3$ pixels gaussian, and shifted in the vertical direction 
for clarity.}
\end{figure*}

We attempted to measure the redshift of all individual objects falling in the slits that revealed a 
discernable continuum or possible emission lines. To obtain an accurate redshift measurement for 
foreground galaxies, cluster members and other bright objects, we applied the IRAF task {\tt xcsao} 
from the Radial Velocity package RVSAO \citep{Kurtz} on all extracted spectra. This procedure 
uses a cross-correlation method based on spectral templates \citep{CWW,Kinney} to estimate 
the redshift and the corresponding redshift error. For the remaining objects in the spectroscopic 
catalog, the redshift measurement is based on the wavelength at the peak of the brightest 
emission line detected. In the latter case, we estimated the redshift error from the spectral 
dispersion.  Additional uncertainties generated by the accuracy of the relative and absolute 
wavelength calibrations, of about 0.8 and 1.5 \AA\ respectively for the LRIS data, were quadratically 
added to the previous estimates to yield the final redshift errors.

A confidence class $C$, ranging from 1 to 4 was assigned to each individual redshift measurement 
according to the prescription of \citet{Lefevre95}: this corresponds to a probability level 
for a correct identification of 50, 75, 95 and 100\%, respectively. A specific value of 9 is used 
when only a single secure spectral feature is seen in emission. The identification of most
$z>2$ objects in the catalog is based on only a single emission line interpreted as Lyman-$\alpha$ 
and a confidence class of 9. However, the constraints provided by the lensing 
configuration in case of multiple images (see Sect. \ref{systems}) enable us to strengthen 
most of these interpretations.

Multicolor photometry was performed for all sources identified in the WFPC/F702W band 
and CFH12k-$R$ band images, using the \textit{SExtractor} software \citep{SExtractor}. 
Total magnitudes and errors were measured on the original images without any resampling or 
convolution. 

The final spectroscopic catalog of lensed galaxies is presented in Table \ref{speccat}.

Figure \ref{field} displays the location of all sources in the catalog and the spectroscopic 
configuration of each dataset. A red circle marks the location of 44 cluster members from 
our sample, for which we measured a spectroscopic redshift. Long-slit and IFU pointings were 
mainly focused on the central region of the cluster (within $\sim$80\arcsec), in the vicinity 
of the critical curves, whereas MOS surveys with LRIS probed a wider field of view of 
$\sim 6\times 8$ arcmin.

The redshift distribution of lensed background galaxies appears to be highly correlated. 
Three sources are located in the $2.63<z<2.69$ range, two sources have $z\sim 0.86$, 
and C1 system has a $z\sim 1.6$ redshift similar to C0 \citet{Smith02b}. 
Even more prominent is a group of 8 sources having $0.62<z<0.64$, which lie predominantly 
at the south of the cluster center, and exhibit similar colors in the composite CFH12k image 
(Figure \ref{field}).

\section{Catalog of Multiply-Imaged Systems}
\label{systems}

\begin{figure*}
\centerline{\mbox{\includegraphics[width=16cm]{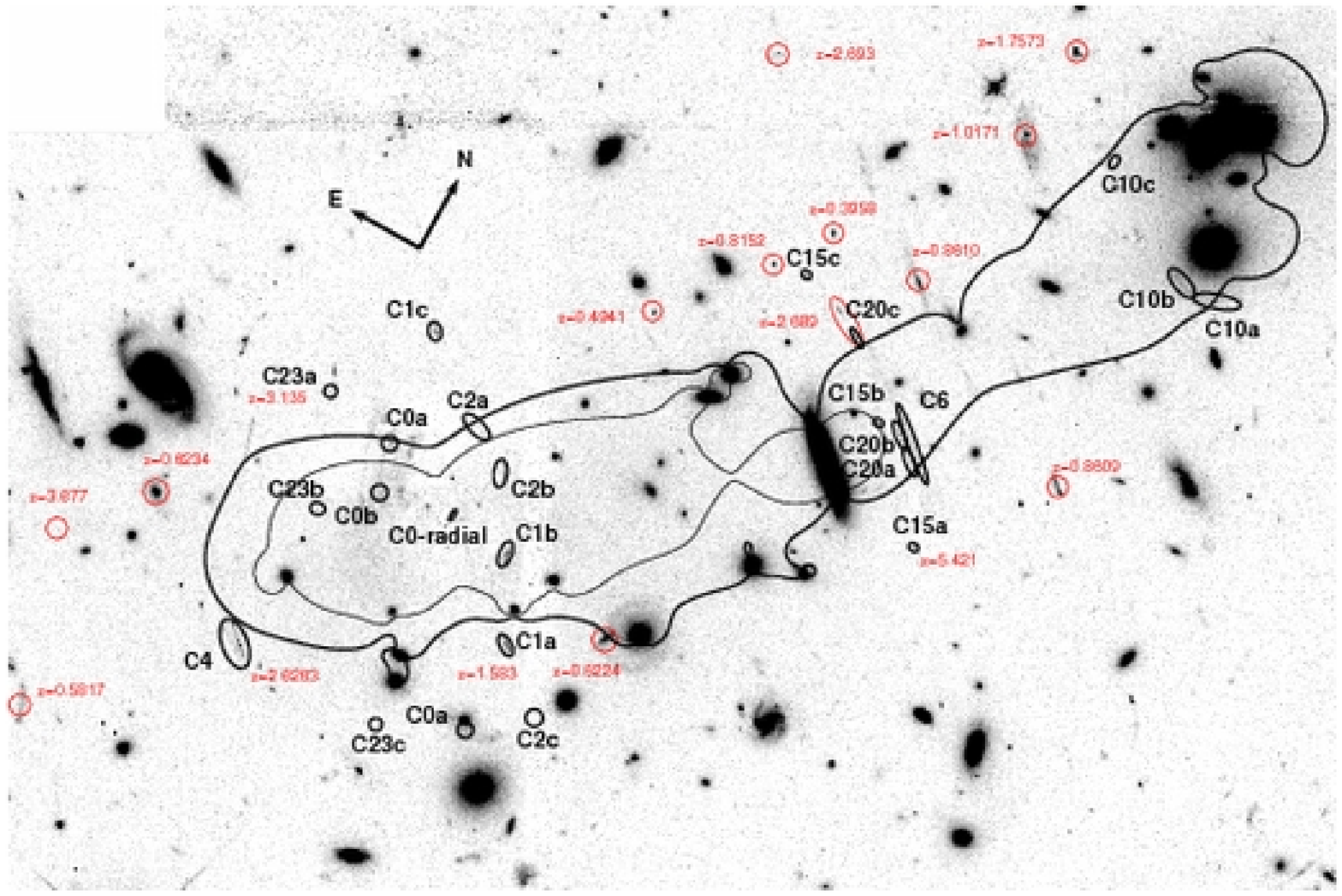}}}

\mbox{\includegraphics[width=9cm]{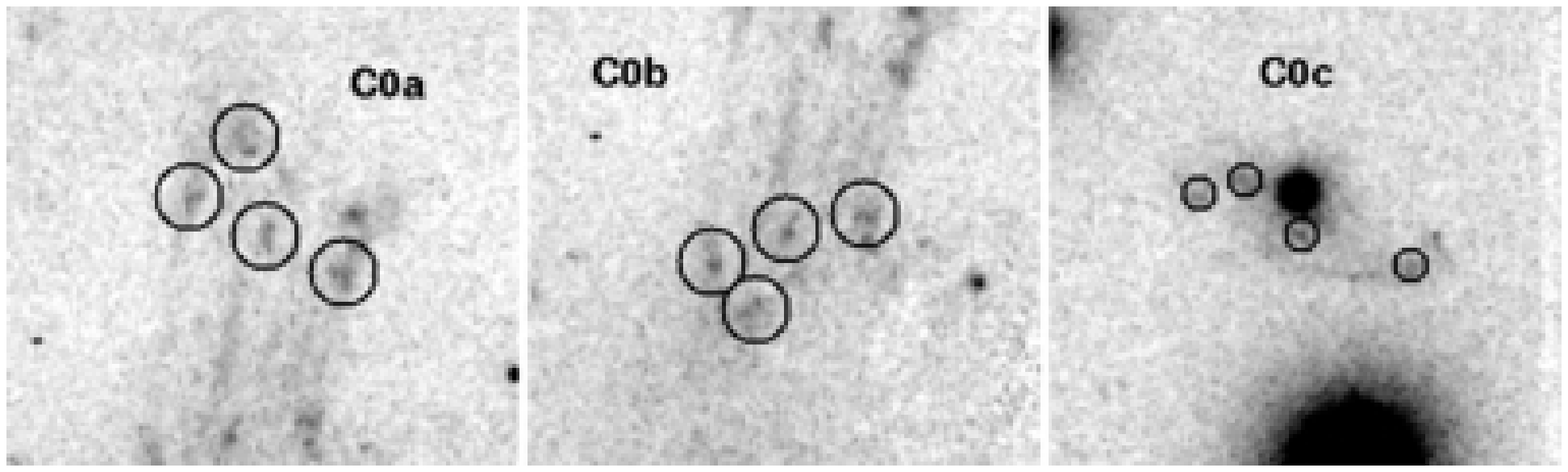}}\mbox{\includegraphics[width=7cm]{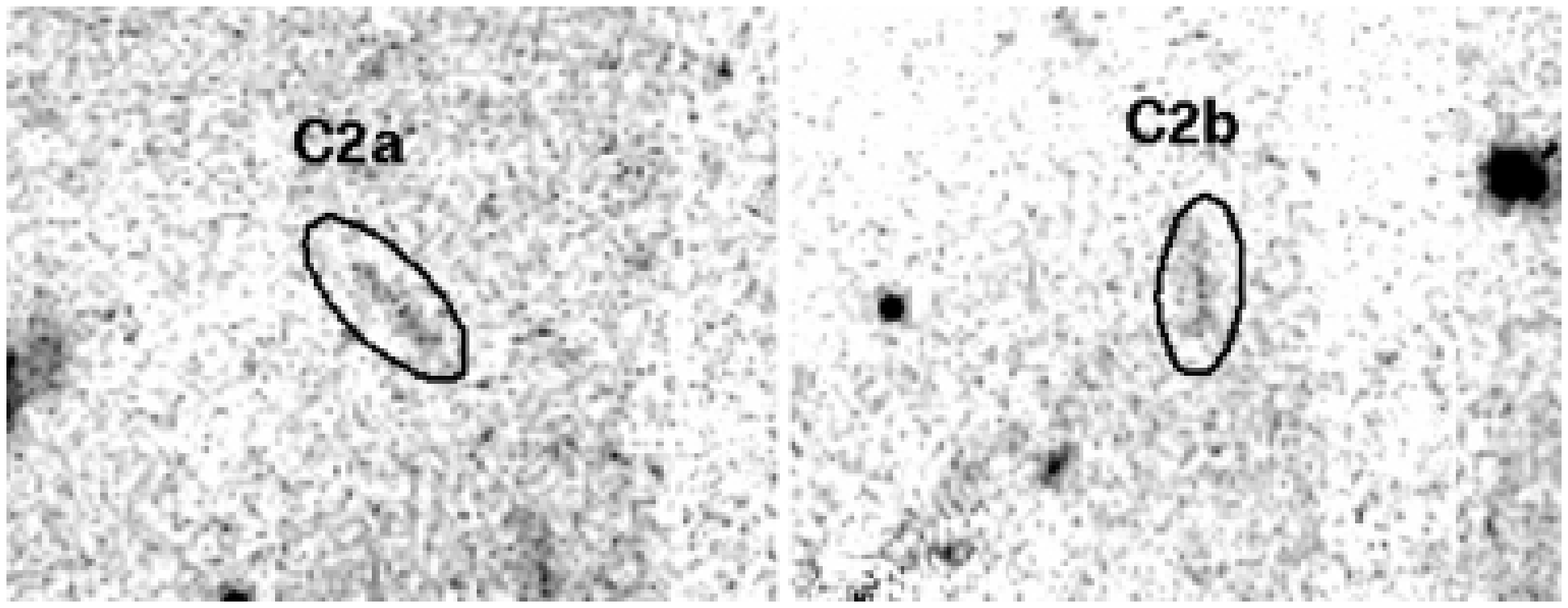}}
\mbox{\includegraphics[width=9cm]{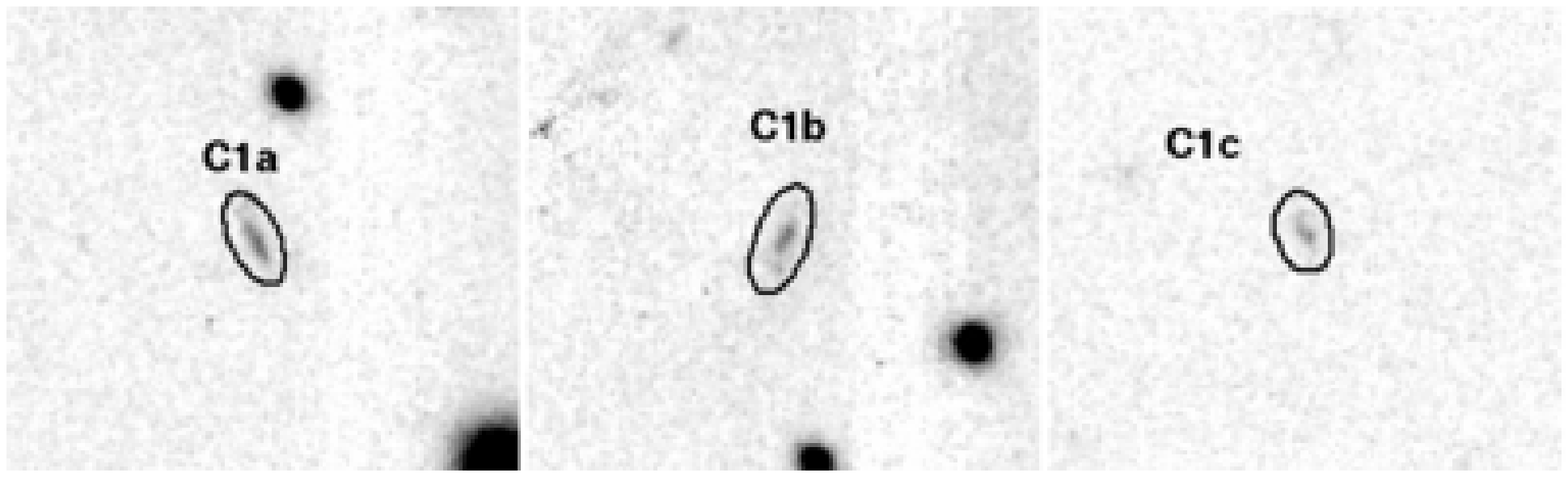}}\mbox{\includegraphics[width=7cm]{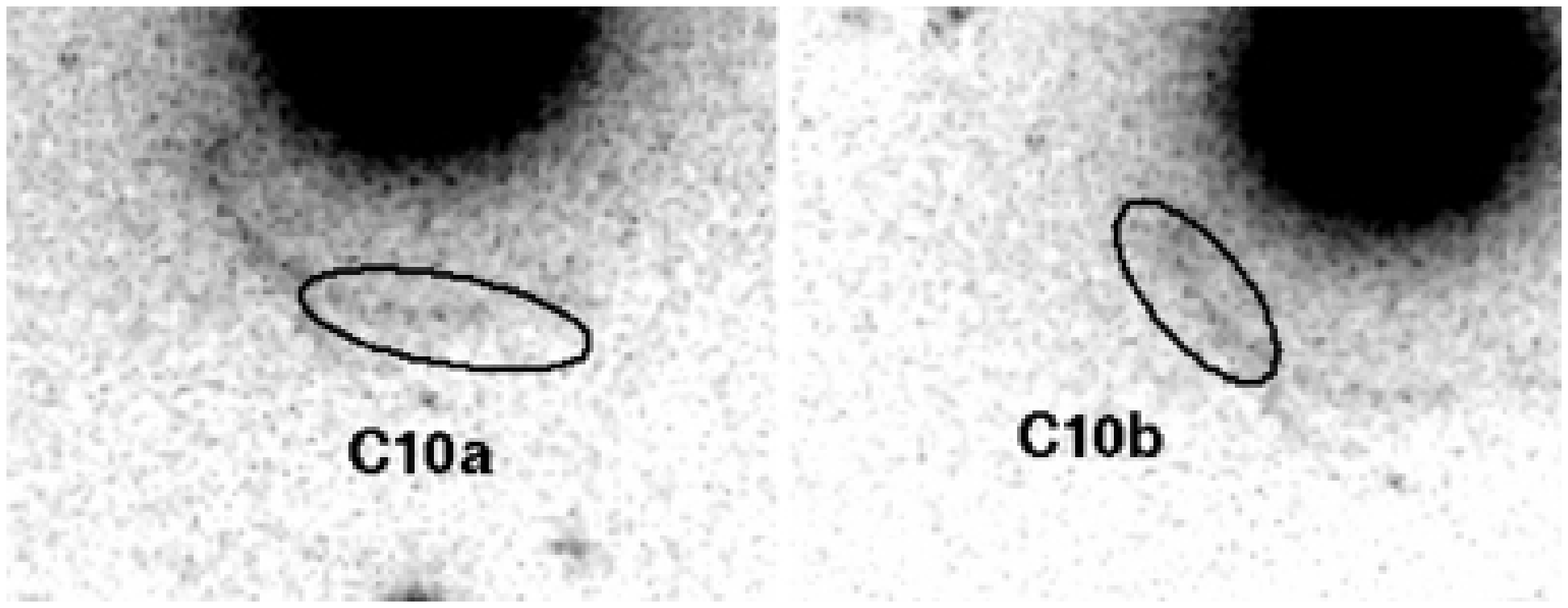}}
\mbox{\includegraphics[width=9cm]{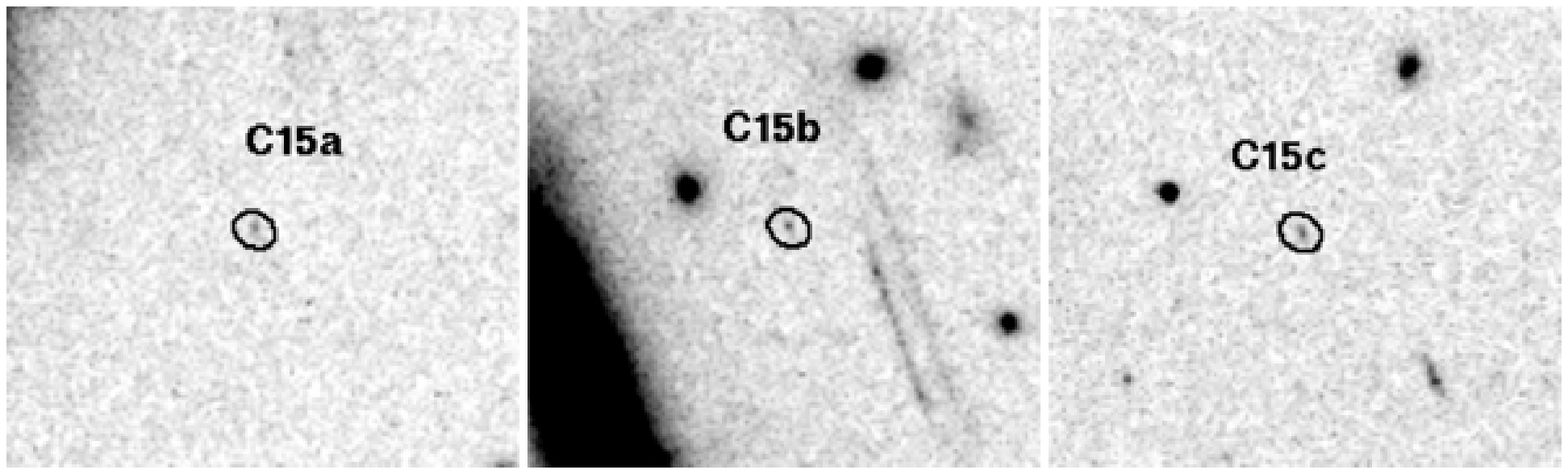}}\mbox{\includegraphics[width=7cm]{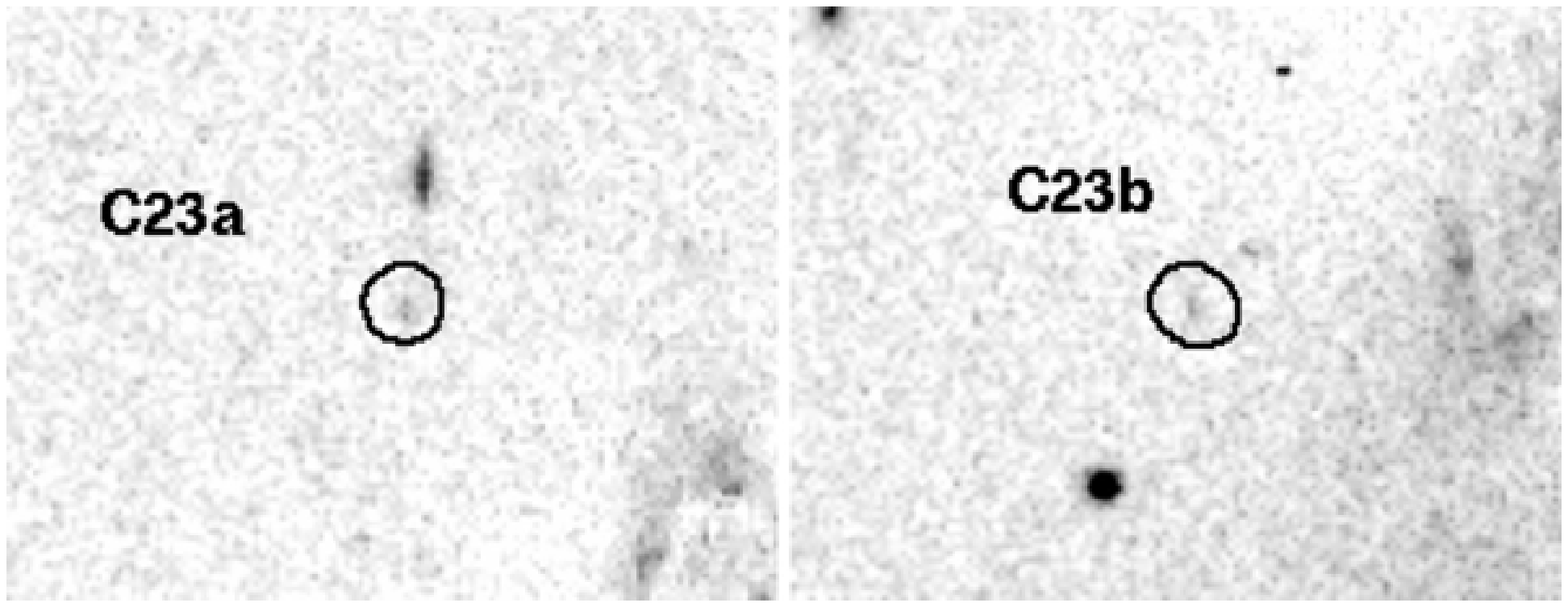}}
\caption{\label{multiple}(Top) Enlarged region of the $R$-F702W image showing the location 
and morphology of the multiple-image systems identified in the field of Abell 68. We subtracted 
the contribution from the brightest cluster galaxy (BCG) to assist in the identification of underlying 
objects, such as the radial arc associated with C0. The thin/thick black curves show the location 
of the critical lines at $z=1.6$ and $z=5.42$ respectively, as inferred from the mass model ($\S$4). We 
overplot in red new spectroscopic redshifts of background sources. (Bottom) Close-up on each 
component of the multiple systems discussed in the text.}
\end{figure*}

In this section we update the catalog of multiply-imaged systems in the field of Abell 68,
taking into account both our new spectroscopic measures and systems without
spectroscopy located on the basis of their geometrical location and similar colors.
For each system, we present a close-up view from the HST-$R$ image in the bottom panels of Figure \ref{multiple}, 
and summarize the position, photometry, shape parameters and magnification in Table \ref{sysprop}. 

\begin{table*}
\begin{tabular}{rllllllll}
\tableline
\tableline
System & RA & DEC & a & b & $\theta$ & $R$ & $z$ & $\mu$ \\
& 00: & 09: & \arcsec & \arcsec & [deg.] & [mag] & & [mag] \\
\tableline
C0 a & 37:07.426 & 09:28.42 &  0.66 &  0.33 & 24.1 & 25.72$\pm0.10$ & 1.6\tablenotemark{a} & 3.14$\pm0.06$\\
   b & 37:07.324 & 09:24.03 &  0.63 &  0.36 &  8.1 & 25.20$\pm0.08$ &     & 2.72$\pm0.06$\\
   c & 37:06.161 & 09:08.84 &  0.42 &  0.30 & 82.3 & 25.84$\pm0.10$ &     & 2.15$\pm0.11$\\
radial & 37:06.870 & 09:25.51 & 0.61 & 0.19 & 350.0 & 26.51$\pm0.30$ &     & 1.75$\pm0.12$ \\
\tableline
C1 a & 37:06.190 & 09:17.42 &  1.38 &  0.48 & 42.7 & 24.02$\pm0.04$ & 1.583 & 2.52$\pm0.06$\\
   b & 37:06.466 & 09:24.80 &  1.14 &  0.63 & 10.1 & 24.19$\pm0.05$ &       & 2.29$\pm0.06$\\
   c & 37:07.515 & 09:39.42 &  0.84 &  0.48 & 41.1 & 24.94$\pm0.07$ &       & 1.62$\pm 0.04$ \\
\tableline
C2 a & 37:07.024 & 09:33.73 &  0.75 &  0.39 & 67.1 & 25.59$\pm0.15$ & $[1.39\pm0.08]$  & 3.09$\pm0.06$ \\
   b & 37:06.724 & 09:31.05 &  0.60 &  0.24 & 38.6 & 25.57$\pm0.16$ &       & 2.80$\pm0.05$\\
 $[c]$& $[$37:05.822$]$ & $[$09:12.94$]$ & &  & & $[26.4]$ &  & 1.83$\pm0.08$ \\
\tableline
C15 a & 37:04.294 & 09:43.41 &  0.42 &  0.21 & 37.3 & 26.00$\pm0.18$ & 5.421& 2.74$\pm0.08$ \\
    b & 37:04.861 & 09:51.79 &  0.27 &  0.21 & 39.6 & 26.52$\pm0.22$ & 5.421& 2.89$\pm0.07$\\
    c & 37:05.691 & 10:00.41 &  0.39 &  0.24 & 41.6 & 26.07$\pm0.17$ &      & 2.69$\pm0.10$\\
\tableline
C20 a & 37:04.560 & 09:49.94 &  0.81 &  0.30 & 44.6 & 25.46$\pm0.11$ &       & 4.9$\pm 0.3$ \\
    b & 37:04.707 & 09:51.85 &  0.96 &  0.30 & 49.2 & 24.83$\pm0.06$ &       & 5.5$\pm0.7$\\
    c & 37:05.231 & 09:57.58 &  0.59 &  0.28 & 59.2 & 26.32$\pm0.16$ & 2.689 & 3.61$\pm 0.09$\\
\tableline
C23 a & 37:07.894 & 09:29.88 &  0.36 &  0.27 & 39.0 & 26.44$\pm0.18$ & 3.135& 2.37$\pm0.07$ \\
    b & 37:07.618 & 09:19.99 &  0.42 &  0.21 & 33.7 & 26.80$\pm0.23$ &      & 2.57$\pm0.06$ \\
  $[c]$ & $[$37:06.634$]$ & $[$09:05.43$]$ &   &    &   & $[27.4]$ &            & 1.73$\pm 0.05$ \\
\tableline
\multicolumn{4}{l}{Possible multiple-image system}\\
C10 a & 37:03.672 & 10:16.75 & 1.0 & 0.25 & 55.0 & 24.56$\pm 0.08$ & [1.11$\pm0.12$] & 3.5$\pm 0.6$\\
    b & 37:03.458 & 10:15.85 & 1.4 & 0.18 & 107.0 & 25.27$\pm 0.13$ &  & 3.7$\pm 2.0$\\
    $[c]$ & $[$37:04.290$]$ & $[$10:23.14$]$ & & & & & & \\
\tableline
\multicolumn{4}{l}{Maximum redshift of single images}\\
C3  & 37:08.203 & 09:22.70 & 1.9 & 0.5 & 108 & & $[<2.7]$ & \\
C5  & 37:06.404 & 09:48.42 & 1.1 & 0.4 & 154 & & $[<2.7]$ & \\
C9  & 37:04.235 & 10:01.67 & 1.7 & 0.3 & 146 & & $[<2.3]$ & \\
C11 & 37:04.789 & 10:03.60 & 1.2 & 0.6 & 152 & & $[<1.9]$ & \\
C13 & 37:04.498 & 10:28.08 & 2.6 & 0.4 & 115 & & $[<1.5]$ & \\
C18 & 37:07.945 & 09:31.83 & 1.2 & 0.4 & 114 & & \multicolumn{2}{l}{No Constraints } \\
\end{tabular}
\caption{\label{sysprop} Properties of the multiple-image systems. In addition to the 6 secured systems, C10 
is included as a potential candidate (see Sect. \ref{riz}). From left to right:
identification, astrometric position (J2000), ellipse shape parameters (a, b, $\theta$), 
$R_{702W}$ magnitude, measured redshift, 
magnification factor (in magnitudes) derived from the mass model. Bracketed values are also predictions 
inferred from the mass model.}
\tablenotetext{a}{\citet{Smith02b}}
\end{table*}

\subsection{Morphological Identification}
\label{riz}

Multiply-imaged systems can be identified via a visual inspection of morphologies in the central 
region of the HST-$R_{702W}$ image (Fig. \ref{multiple}) , examined in combination with the broad-band photometric catalogs. Our 
starting point is a detailed study of candidates identified to $\mu_{702}\sim$ 25 mag arcsec$^{-2}$
presented by S05. This work revealed 3 main systems (C0a,b,c; C1a,b,c; C2a,b) as well as 
20 other possible multiply-imaged candidates (C3 to C22). We adopt the same nomenclature, 
extending to include the new systems presented here.

The three images identified as C15, C16 and C17 have been selected as $R$-band dropouts 
on the basis of color-color diagrams combining $R$, $I$ and $z$-band filters. Indeed, they are 
very faint in the HST image ($R_{702}\sim 25.7$), undetected at shorter wavelengths in the 
$B$ or $R$ band with CFH12k with a combination of red $(R-I)_\mathrm{AB}\sim 1.7$ and 
blue $(I-z)_\mathrm{AB}\sim -0.5$ colors, as measured with aperture photometry on the 
seeing-matched images. Such a spectral energy distribution (SED) is characteristic of 
galaxies in the redshift range $5\lesssim z\lesssim 6$, where the Lyman-$\alpha$ break 
in the spectral continuum occurs between the $R$ and $I$ bands. The preliminary mass 
model ($\S$4) shows that the location of these three images is compatible with a 
multiply-imaged source, thereby strengthening the high redshift interpretation. We designate 
this system by C15 (with three images C15a,b,c that correspond to C15/16/17) and 
confirmed the components C15a and C15b lie at the same redshift using LRIS spectroscopy 
(see Section \ref{redshifts}).

The arclet identified as C20 was serendipitously covered during the same LRIS long-slit 
observation.  We interpret the extended emission seen in the spectrum (Sect. \ref{redshifts}) 
as Lyman-$\alpha$ at $z=2.689$. This is also supported by the mass model, which predicts 2 
detected counter-images with similar optical colors. We refer to this system as C20a,b,c.

Next to C20a and C20b images, a fainter extended arc C6 was identified by S05 as a possible 
multiply-imaged system at a similar redshift. This arc can be split into 3 components C6a,b,c.

Finally, we uncovered a new lensed image, C23a, close to the cluster center during the critical 
line survey using LRIS \citep{Santos}. Its spectrum is consistent with a high redshift source 
dominated by its Lyman-$\alpha$ emission. Again, the cluster mass model predicts the position 
of a second faint component for this system (C23b), detected on the HST image. 

\subsection{Redshift Constraints}
\label{redshifts}

We now present the new spectroscopic redshifts obtained for 6 multiply-imaged candidates, as 
well as constraints implied for the remaining multiply-imaged systems C2 and C10, from the 
updated mass model (Sect. \ref{mmodel}). 

\begin{itemize}
\item{C0 source}

This source is a strongly-lensed ($\mu\sim 3.2$ mag.) multiply-imaged system with three components, 
discovered during a survey for Extremely Red Objects (EROs) in the fields of 10 massive galaxy 
cluster lenses at $z\sim0.2$ \citep{Smith02a}. A redshift measurement on the brightest image, C0a, 
was presented in \citet{Smith02b}. Based on the 4000 \AA-break identification at $\lambda=
1.04\pm 0.01 \mu$m, the redshift is $z=1.60\pm 0.03$. This image was also included in two LRIS 
masks, but we failed to detect strong spectral features in the wavelength range 5500-9200 \AA\ . 

Because of the clear symmetry of images C0a and C0b with respect to the critical line, it is possible 
to isolate 4 bright knots in each image $a$, $b$ and $c$ and use the corresponding 12 images to 
constrain the mass model. Furthermore, part of the south-west knot of C0a is located within the 
radial caustic line in the source plane. We are able to detect a very faint radial arc predicted by the 
mass model (labeled C0-radial on Fig. \ref{multiple}) after removing the light from the BCG.  

\item{C1 source}

The brightest component C1a of this system was observed during our NIRSPEC critical lines
survey (see \citet{Stark} and Stark et al.\,(2006b), in press). Three strong emission lines 
were identified as $[O III] {5007}$, $[O III] {4959}$ and H$\beta$ (Figure \ref{C1a}, left), which 
unambiguously give a redshift $z=1.583$. Optical spectroscopy of component C1c,
 included in one of our LRIS masks, could not identify any strong $[O_{II}]$ emission for this source, 
with a 3$\,\sigma$ upper limit of $50$ \AA\ for the equivalent width in rest-frame.

\item{C4 source}

Although not multiply-imaged, this blue arc has very strong Lyman-$\alpha$ emission 
corresponding to $z\sim 2.63$. An extended Lyman-$\alpha$ blob has also been detected 
in the IFU data (see Sect. \ref{mapping} and Figure \ref{multiple}). The mass model gives 
a very high magnification factor of $\mu \sim 4.0$ mag., because the source 
is very close to a cusp.

\item{C6 and C20 sources}

We identify a spatially-extended emission line in the LRIS spectrum at 
$\lambda=4485$ \AA\ , surrounding the C20c image (Fig. \ref{C1a}, middle).  The lack 
of other strong emission lines in the optical range and the predicted position for two other 
components for C20 from the mass model confirms this to be Lyman-$\alpha$ at $z=2.689$.  
In addition, this redshift is close to the Lyman-$\alpha$ blob associated with C4, and 
we identify a similar strong emission line at a similar redshift, $z=2.693$, for the nearby 
source C25 ($\sim8$\arcsec\ away in the source plane). 

The location and color of the lower surface brightness arc, C6, close to images C20a and 
C20b, is compatible with three merging images forming a single giant arc at a similar 
redshift (Fig. \ref{C1a}, middle). This is further evidence of a possible group of galaxies 
at $z\sim 2.6$ including sources C6, C20 and C25.

\begin{figure*}
\mbox{\includegraphics[width=4.5cm]{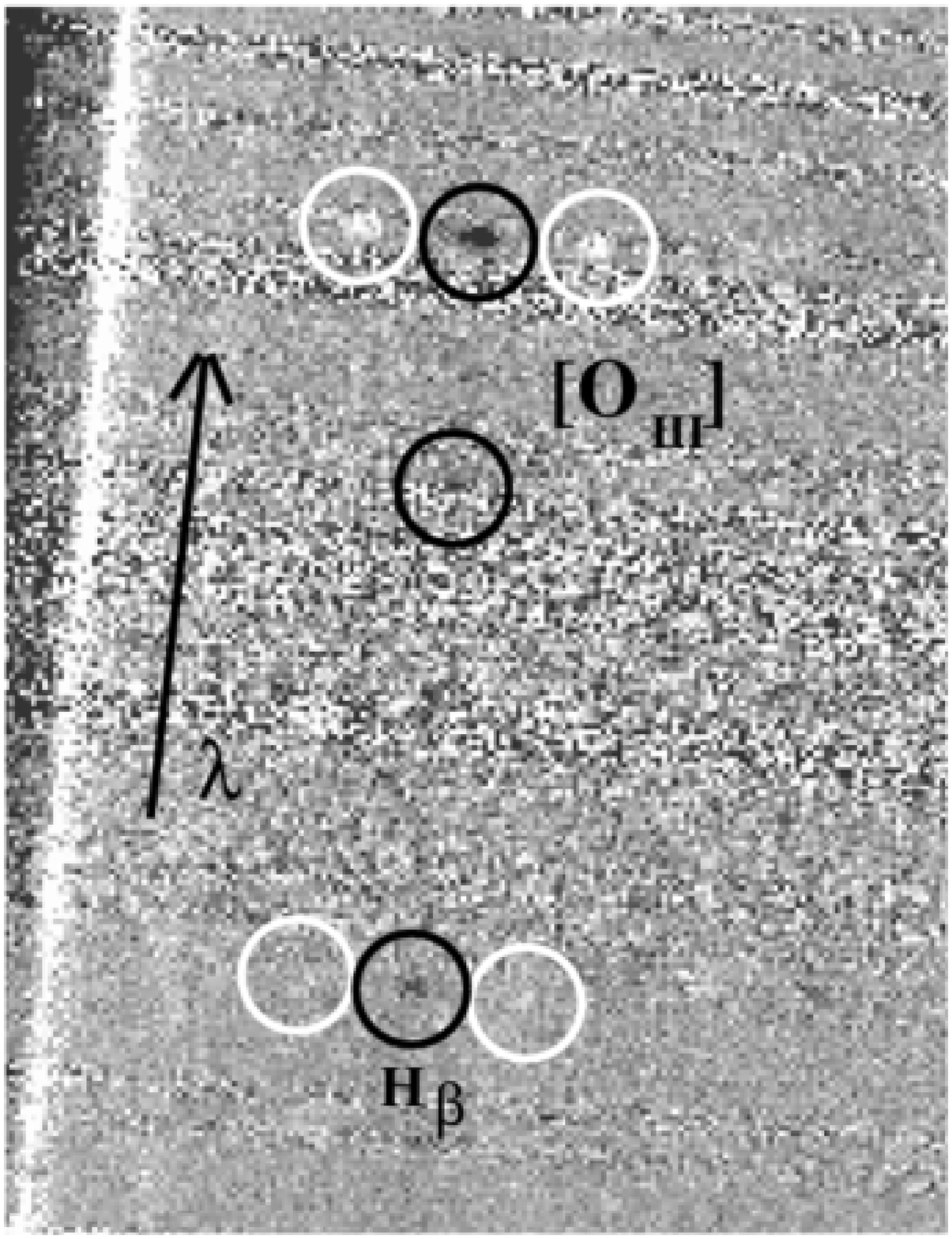}}
\hspace{.5cm}
\mbox{\includegraphics[width=5.2cm]{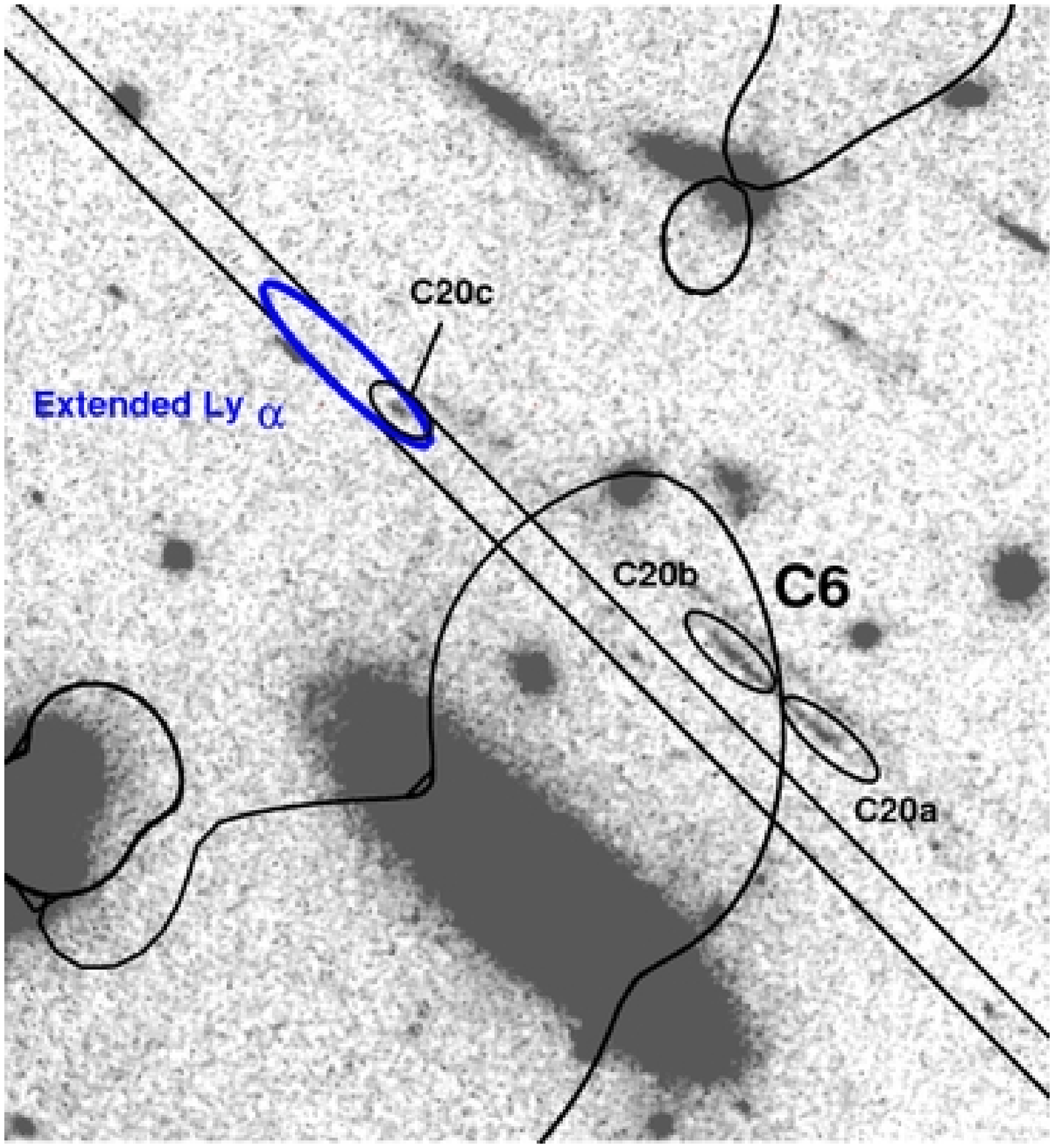}}
\hspace{.5cm}
\includegraphics[width=4.5cm]{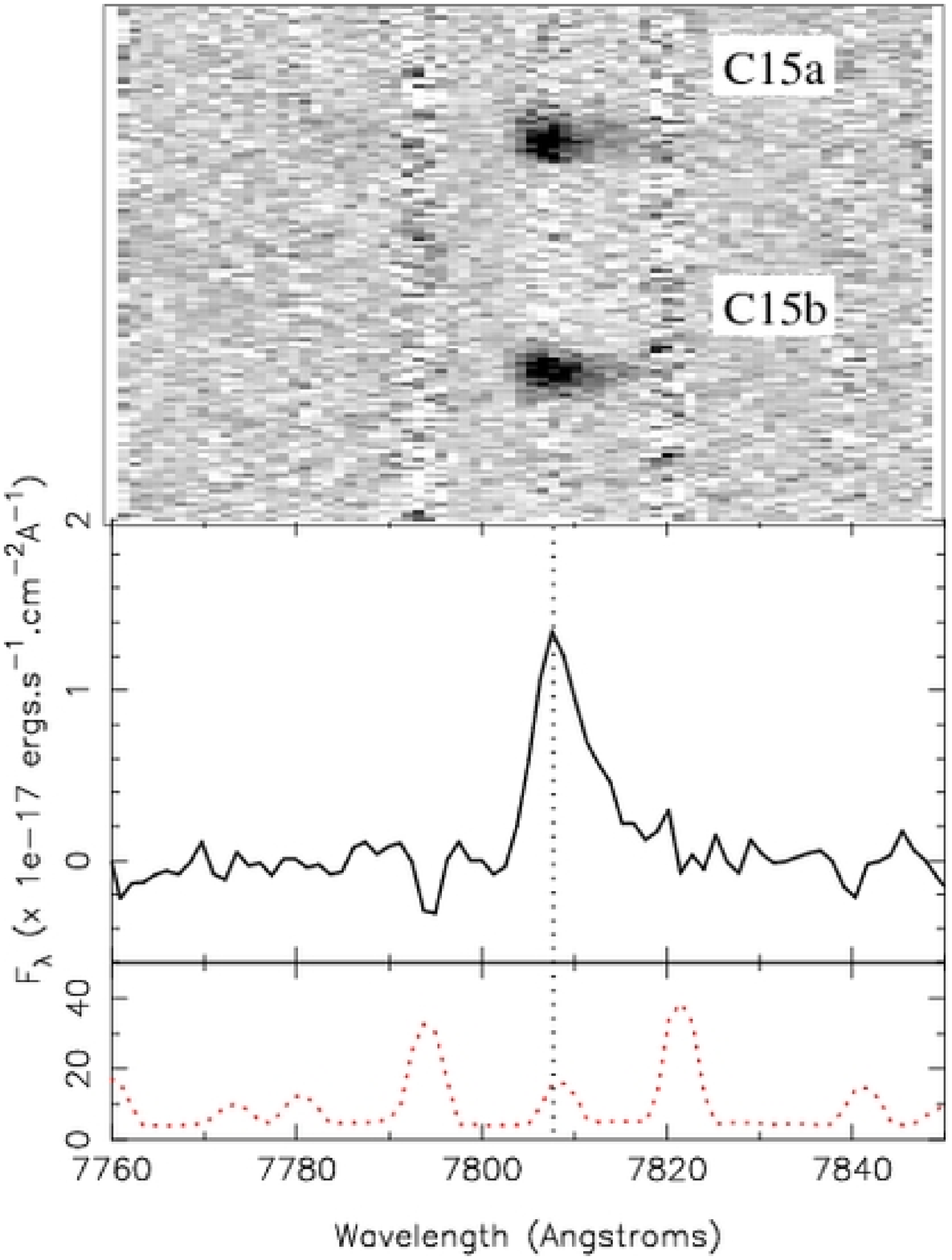}
\caption{\label{C1a}(Left) NIRSPEC bidimensional spectrum of component C1a, showing the 
detection of 3 strong features identified as $[O_{III}]$ and H$\beta$ emission lines at $z=1.583$, 
both in positive (black) or negative (white), due to dithering offsets used.
(Middle) Location of the LRIS slit and extended Lyman-$\alpha$ emission identified around 
image C20c (thick ellipse). Two brighter components C20a and C20b are predicted by the 
mass model, in agreement with the location of the critical line at $z=2.68$ (solid curve). The 
shape and colors of the adjacent giant arc C6 are compatible with a similar redshift, suggesting 
a physical connection between these two sources. (Right) Upper panel: close-up of the 2-D 
sky-subtracted LRIS spectrum, showing both strong emission lines seen at the position of 
C15a and C15b. Lower panel: average extracted spectrum of C15, with the same wavelength 
range, revealing the clear asymmetrical shape of the spectral line. The relative sky background 
level is presented as a dotted curve.}
\end{figure*}

\item{C15 source}

A strong asymmetric emission line is clearly seen on the LRIS spectrum at the position of C15a and 
C15b (Figure \ref{C1a}, top-right), with a central peak at 7808 \AA . In both cases, a faint continuum 
is detected on the red side of the emission line, with a flux $\sim 4.4\pm 1.5\times10^{-19}$ ergs\, 
s$^{-1}$\ cm$^{-2}$\ \AA$^{-1}$. We interpret the emission line as Lyman-$\alpha$ at $z=5.421$. 
This redshift could be slightly overestimated due to an unknown amount of self-absorption in the 
Lyman-$\alpha$ emission line and the absence of other spectral features. By averaging the 
extracted spectra of both images (Figure \ref{C1a}, top-right), the integrated flux measured within 
the emission line, of $9.7\pm0.8\times10^{-17}$ ergs\, s$^{-1}$\ cm$^{-2}$, corresponds to a 
rest-frame equivalent width of $\sim 35$ \AA . We measure a Full-Width Half-Maximum of $\sim 
7$ \AA\ when fitting the average emission line with a gaussian profile. The strong Lyman-$\alpha$ 
emission line of this source seems to dominate the overall $I$-band flux.

\item{C23 source}

The identification of the pair of images C23a and C23b, predicted by the lensing model at $z=3.1$, 
strengthens the Lyman-$\alpha$ interpretation for the single emission line seen in the spectrum of 
C23a at $\lambda=5028$ \AA. A fainter counterimage, C23c, predicted by the mass model, 
lies beyond the detection limit of the HST image. 

\item{C2 source}

This system is composed of two bright symmetrical images, identified close to the cluster center. 
Using the mass model, we predict a redshift of $z\sim 1.4$ for this source, with a magnification factor 
of $\mu\sim2.9$ mag. for C2a and C2b. A less magnified ($\mu \sim 1.8$ mag.) counter-image C2c is also 
predicted by the model, but is not detected in the less-sensitive region of the HST image, at the junction between 2 WF chips.

\item{C10 source}

A faint curved arc near a bright cluster member was identified by S05 as a multiple imaged candidate C10. 
The optical colors of this blue arc are indeed compatible with a source at $z<2.5$. It is probably formed by two images merging 
on the critical line; its shape is in agreement with predictions from the mass model. However, the low 
surface brightness of this arc and the lack of spectroscopic information strongly limit this interpretation, making 
it more uncertain than the systems previously mentioned. 
Therefore, we only include this source in Table \ref{sysprop} as a possible additional multiple imaged system. The mass model predicts a redshift of 
 $\sim 1.1$ which is in very good agreement with the photometric redshift estimate of 
 $z_{phot}=1.15\pm0.05$ derived from the broad band colors using the \textit{Hyperz} photometric 
 redshift code \citep{Hyperz}. A fainter counter-image (C10c) is predicted by the mass model, but 
 remains undetected in the HST image. 

\end{itemize}

\section{Mass model}
\label{mmodel}
\subsection{Modeling method}

Using the new redshift measurements and the identification of further multiply-imaged systems, we 
are now in a position to considerably improve the precision of the mass model presented by S05.
In doing so, we will maintain the Pseudo-Isothermal Elliptical Mass Distribution model 
(PIEMD, Kassiola \& Kovner 1993) adopted by S05 to infer the dark matter mass distribution. This 
parametric method has been used for modeling galaxy clusters as well as individual galaxies 
\citep{Covone, Natarajan}. It assumes each dark matter clump can be parameterized by a central
position, ellipticity $a/b$, position angle $\theta$, central velocity dispersion $\sigma_0$ and two 
characteristic radii $r_{core}$ and $r_{cut}$. The total mass of this profile is proportional to $r_{cut}\ \sigma_0^{2}$. A more detailed discussion of the validity of this approach, in contrast to alternatives,
is given by \citet{Limousin07}.

The cluster galaxy population is incorporated into the lens model as galaxy-scale perturbations 
to the cluster potential, assuming a scaling relation $M/L=\mathrm{Const}$ \citep[][S05]{Brainerd}
for all the $r_{core}$, $r_{cut}$ and $\sigma_0$. This is motivated by the similar \citet{FJ} scaling 
relation observed in elliptical galaxies. Following the same procedure as \citet{Limousin07}, we 
keep the $r^{*}_{cut}$ and $\sigma^{*}_0$ values corresponding to a $L^{*}$ elliptical 
galaxy as free parameters, while keeping $r^{*}_{core}$ at 0.15 kpc. We select cluster galaxies 
in the field of Abell 68 by plotting the characteristic cluster red sequences ($B$-$R_{702}$) and 
($R_{702}$-$K$) in two color-magnitude diagrams, keeping the objects pertaining to both red 
sequences. This reduces the photometric catalog used by S05, containing 69 galaxies 
with $K$ band photometry, to 47 cluster members. This method is more efficient than a single 
red sequence selection as we did not select any of the spectroscopically confirmed 
background or foreground galaxies. As described in S05, the $K$-band photometry was obtained 
by \citet{Balogh} using GIM2D \citep{Simard} to fit the surface brightness profiles of the cluster 
galaxies. This method gives more accurate results than one could obtain with SExtractor, because 
SExtractor usually overestimate the local background around the brightest and most extended galaxies.\\

In a first attempt at modeling the lens, we adopted the parameters given by S05, who included two 
dark matter halos. We found any attempt to reconstruct the mass distribution using a single clump 
was unable to reproduce the multiply-imaged systems accurately, confirming the strong bimodality 
of this cluster. By incorporating our spectroscopically-confirmed multiply-imaged systems, 
a reoptimization is necessary. To do this, we use the new bayesian optimization method provided 
by the \textit{Lenstool} software\footnote{For more information and to download the latest version 
of the code, see {\tt http://www.oamp.fr/cosmology/lenstool}}\citep[][Jullo et al. in prep.]{Kneib93}
so that we can derive error estimates for each optimized parameter. 
The software optimizes the locations of each system in the source plane, based on the following 
 $\chi^2$ estimator, which defines the goodness of the fit:
\begin{equation}
\chi^{2}=\sum_{i}\chi_i^{2}
\end{equation}
where $\chi_i^{2}$ is the same estimator for a given multiply-imaged system $i$, constructed by comparing the 
predicted positions of the N observed images in the source plane ($x_{i,j},y_{i,j}$) to their barycenter 
($x_i^B,y_i^B$):
\begin{equation}
\chi_i^{2}=\frac{1}{N}\sum_{j=1..N}\frac{(x_{i,j}-x_i^B)^{2}+(y_{i,j}-y_i^{B})^{2}}{\sigma_{pos}^{2}}
\end{equation}
$\sigma_{pos}$ is the uncertainty in measuring the position of each image in the source plane. We 
use a typical value of $\sigma_I=0.2\arcsec$ for the same uncertainty in the image plane, and relate 
it to $\sigma_{pos}$ with the amplification $A$: $\sigma_I^{2}=A\ \sigma_{pos}^{2}$.

For its important influence on the position of systems C15 and C20, we choose to optimize the velocity dispersion and the cut radius of the third brightest cluster galaxy. We also optimize the same parameters 
for the BCG and the galaxy \#106 (adjacent to image C0c), in order to match the location of 
all images from the C0 system. 

The associated reduced $\chi^2$ is $\sim1.7$, with 21 degrees of freedom, and the astrometric 
error on the position of the predicted multiple images is $0.37\arcsec$ in the image plane. The 
new parameters of the mass model are presented in Table \ref{parameters}. The optimized values 
are slightly higher than the ones we obtain by using only their luminosity as a parameter for the 
gravitational potential. However, those values are not sufficiently high to propose a third dark 
matter clump. As additional confirmation for the quality of the mass model, we also checked that all 
spectroscopically-confirmed background sources for which we did not identify multiple-image 
systems were predicted to be singly-imaged. 

In comparison with the results from S05, we have refined the mass model by increasing the number 
of constrained parameters from 11 to 21, while keeping a similar reduced $\chi^2$ value. In particular, 
we optimized the location of the second dark matter clump, where we have the main differences in the 
confirmed multiply imaged systems, and constrained individual parameters for two particular galaxies, 
in complement to the BCG.

\begin{table*}
\begin{tabular}{llllllll}
\tableline
\tableline
Mass & $\Delta$R.A. & $\Delta$Dec & $a/b$ & $\theta$  &  $r_{core}$ & $r_{cut}$  & $\sigma_0$ \\
component & (arcsec) & (arcsec) &  & (deg) & (kpc) & (kpc) & ($\mathrm{km\,s}^{-1}$) \\
\tableline
Cluster\#1 & -1.5$\pm0.4$ & 0.2$\pm0.3$ & 1.8$\pm1.0$ & 125.9$\pm0.6$ & 87.9$\pm6.0$ & 1239$\pm471$ & 908$\pm37$\\
Cluster\#2 & -48.4$\pm1.6$ & 63.2$\pm2.2$ & [1.0] & [0.] & 65.1$\pm14.3$ & 1350$\pm281$ & 757$\pm57$\\
BCG       & [0.]    & [0.]   &  [1.3]  & [122.5]  &  [0.25]  &  83$\pm45$ & 266$\pm5$ \\
\#3       & [-27.5] & [22.] & [2.6] & [53.9] &  [0.043]  &  150$\pm34$ & 179$\pm7$ \\
\#106     & [-10.0] & [-14.5] & [1.2] & [0.] &  [0.013]  &  188$\pm63$ & 63$\pm7$ \\
$L^*$ elliptical galaxy & -- & -- & -- & -- & 0.15 & 18$\pm5$ & 179$\pm13$\\
\tableline
\end{tabular}
\caption{\label{parameters} Most likely parameters of the lens model with $1\sigma$ error bars. From left to right: identification, astrometric position relative to the Brightest Cluster Galaxy, PIEMD parameters. $\theta$ orientation is increasing from north through east. Bracketed values are not optimized.}
\end{table*}

\subsection{Results from the Mass Model}

Our improved mass model of Abell 68 enables us to compute a \textit{lensing} mass of the cluster 
by integrating the derived surface mass density within a given projected distance $R$. Within a physical 
radius $R<500$ kpc, we obtain a value $M_{lensing}=5.31\pm0.17\times 10^{14}\ M_{\odot}$. This
is somewhat higher than the value of 4.4$\pm0.1\times 10^{14}\ M_{\odot}$ derived by S05. By 
comparing both models using the individual parameters given in table \ref{parameters}, this difference 
arises mainly because of the higher mass of the second dark matter clump, for which we derive higher 
$\sigma_0$ and $r_{cut}$ values compared to S05. The remaining model parameters (the location, 
ellipticity and orientation) lie within 3$\sigma$. We argue that the parameters of the secondary clump 
are better constrained in our model, given the identification of new multiple images to the NW of the 
cluster. The difference with S05 is further revealed by computing the ratio $M_{cen}/M_{tot}$ between the
 mass of the central component (with sole contributions from the main clump and the BCG) and the 
total mass within 500 kpc. We find $M_{cen}/M_{tot}=0.56\pm0.05$, lower than the value of 0.68 
derived by S05. This strengthens the bimodal nature of Abell 68. Our parameters for the individual 
contributing galaxies are quite similar to S05, except for the low $r^{*}_{cut}$ value, indicating that 
the haloes have a smaller spatial extent.

Within the uncertainties, we are now able to predict the location and expected fluxes of the 
counter-images for the observed multiple systems C2, C10 and C23. Furthermore, the model 
gives us an estimate of the redshift for the multiple systems C2 and C10 which do not yet have 
spectroscopic measurements. These predictions are summarized in Table \ref{sysprop} as 
bracketed values. 

In the process of the bayesian optimization, the software computes the magnification factors and the 
related error estimates, for each source included in the spectroscopic catalog 
or the multiple images catalog. These values are reported in Table \ref{speccat} and \ref{sysprop},
 respectively.

Apart from the multiple-image systems described before, all other background sources 
in our spectroscopic catalog (C4, C7, C8, C12, C14, C24, C25, C26, C27) are predicted to be 
singly-imaged. This is compatible with our morphological data. For other sources that do not 
show multiple images (C3, C5, C9, C11, C13, C18), we use the mass model to predict their 
maximum redshift $z_{max}$. As the radius of the critical line increases with source redshift $z_s$, 
a multiple image is expected if $z_s>z_{max}$. These $z_{max}$ values are summarized in 
Table \ref{sysprop} and are consistent with the observed optical colors. Image C18 is 
predicted to be a single image at any redshift.

\section{Physical Properties of Faint Lyman-$\alpha$ Emitters}

Via our spectroscopic campaign and improved mass model, we are now in a position to 
explore the physical properties and luminosity distribution of a large sample of intrinsically
faint $2<z<6$ sources. We recognize that the volumes sampled may be unrepresentative but that our survey illustrates, by example, the promise of more extensive surveys that may soon
be possible with larger cluster samples.

First we selected from our spectroscopic catalog a subsample of seven high redshift sources 
($z>1.5$), which all have clearly detected Lyman-$\alpha$ emission lines. The measured 
rest-frame equivalent widths $W$ exceed 2 \AA, and the majority have $W\gtrsim 10$ \AA\ .
By contrast, typical LAEs selected within narrow-band surveys have $W\gtrsim 30$ \AA\ . 
The strong magnification thus gives us insight into the physical properties of these faint LAEs, 
including the star formation rate, the stellar mass and the physical scale corresponding to the 
star-forming regions. These physical values are derived and summarized in Table \ref{physics}.\\

\subsection{Star Formation Rates}

We explore two ways of determining the instantaneous star formation rate ($\mathit{SFR}$) for 
a high redshift LAE, both using the calibration from \citet{Kennicutt} who assumed a 
\citet{Salpeter} initial mass function with mass limits 0.1 and 100 M$_{\odot}$. The first calibration 
is based on the UV continuum luminosity $L_{1500}$, at 1500 \AA\ in rest-frame, with the 
following relationship:

\begin{equation}
     \mathit{SFR}_{\mathrm{UV}} (M_\odot\,\mathrm{yr}^{-1})=
     1.05\times 10^{-40}\,L_{1500}\,(\mathrm{ergs\,s^{-1}\,{\AA}^{-1}})
\end{equation}

We estimated the individual $L_{1500}$ values in this sample through the broad-band photometric measurements.\\

The second calibration adopted from \citet{Kennicutt}  is based on the intrinsic luminosity within the 
Lyman-$\alpha$ emission line, assuming  no extinction and case B recombination (Brocklehurst 1971) :

\begin{equation}
    \mathit{SFR}_{\mathrm{Ly}\,\alpha}(M_\odot\,\mathrm{yr}^{-1})=
    9.1\times 10^{-43}\ L_{\mathrm{Ly}\,\alpha} (\mathrm{ergs\,s^{-1}})
\end{equation}

The corresponding $\mathit{SFR}$ estimates are given in Table \ref{physics}. 

In most cases, we find good agreement between these two estimates, with a trend for 
$\mathit{SFR}_{\mathrm{Ly},\alpha}$ to be lower than $\mathit{SFR}_\mathrm{UV}$, with a mean
ratio $\mathit{SFR}_\mathrm{UV}/\mathit{SFR}_{\mathrm{Ly}\,\alpha}\sim1-4$. We interpret this 
difference as due to the specific properties of Lyman-$\alpha$ emission, which usually show  
some self-absorption or dust extinction. The ratio above is quite similar to that, $\sim 3$, typically found 
in high-redshift Lyman-$\alpha$ samples \citep{Santos,Ajiki03}, as well as for the most distant 
galaxies at $z\sim 6.5$ \citep{Hu02,Kodaira}. We notice two exceptions for C27 and C25; these 
show much higher $\mathit{SFR}$ ratios ($\sim 50$). Since C27 shows a double-nucleus, its UV 
emission may be coming from an AGN. For C25, we estimate a 0.6\arcsec-offset between the 
center of the star-forming region and the center of the LRIS slit (originally aligned on C15a and 
C15b). Conceivably we missed the majority of the light  coming from this object, which may explain 
the discrepancy.

\subsection{Stellar Masses}
\label{smass}

Next we combine the constraints from multiband photometry and spectroscopic redshifts to derive 
estimates of the stellar mass associated with each source. A key question is whether LAEs
are being seen at a special stage in their evolution, for example with a high star formation
rate but low stellar mass. In this calculation, prior to SED fitting, it is important to remove the  
contribution of the Lyman-$\alpha$ flux from the photometric measurements.

We derive the stellar mass using the Bayesian stellar mass code developed by \citet{Bundy},
which compares the SED of each object with a grid of synthetic SEDs from \citet{BC03}, assuming 
a \citet{Chabrier} Initial Mass Function. The star-formation history is parametrized as an 
exponential decaying burst $\mathit{SFR}\propto \exp{-t/\tau}$. In addition, the code gives an 
error estimate for each stellar mass, combining photometric errors and degeneracies in the 
model parameter space (age, reddening, metallicity, star-formation history). At  $2\lesssim z\lesssim 5.5$, 
photometric measurements or upper limits in the 7 filters from Table \ref{images} cover the restframe UV and optical
wavelengths, which is the main limiting factor in deriving stellar masses. Depending on the number of 
photometric bands (2 to 7) where a LAE is detected, the typical errors arising from the degeneracies in 
model parameters cover the range 0.1 to 0.5 in $\mathrm{log}_{10}(M_*)$. It was not possible to derive a stellar mass for C26, 
because it is only detected in the $R_{F702W}$ filter.

Stellar masses, corrected for magnification, are reported in Table \ref{physics}. We also compute  
the inverse of the specific star formation rate (star formation rate per unit stellar mass, \citet[e.g.][]{Brinchmann}) based on the $\mathit{SFR}_{\mathrm UV}$ estimate. This gives an indication on which timescale the star formation is taking place in each object. We find typical 
values of 500 Myr to 1 Gyr, suggesting that higher star formation may have occured in these LAEs during the past, unless they were 
formed very early.

\subsection{Physical scales}
\label{phys_scale}

In addition to brightening the observed flux of background sources, the magnification also 
stretches the angular sizes of the lensed images. This affects all the object shape parameters 
($a$, $b$, $\theta$), increasing the observed solid angle by the same factor $\mu$. We are 
thus able to probe a physical scale $R$ in the source plane, with:

\begin{equation}
R=\sqrt{\frac{a\ b}{\mu}}\ D_{\mathrm{A}}(z)
\end{equation}

\noindent where $D_A(z)$ is the angular distance for a source at $z$.

For most of the high redshift sources which are not resolved (along one direction) in the HST image, 
this value of $R$ is an upper limit (Table \ref{physics}). In 
most cases, we find that we are able to probe high redshift sources at sub-kiloparsec scales. 
Another interesting physical property we derive is the intrinsic surface density of 
star formation ($\Sigma=\mathit{SFR}_\mathrm{UV}/(\pi\ R^{2})$),  which is a lower limit when 
sources are unresolved.

\begin{table*}
\footnotesize
\begin{tabular}{lllllllll}
\tableline
\tableline
                        & C27          & C4          & C20c        & C25         & C23a         & C26          & C15ab       & Ellis et al.\\
\tableline
$z$                     & 1.75         & 2.63        & 2.69        & 2.69        & 3.13         & 3.68         & 5.42        & 5.58 \\
$\mu$ (mags)            & 1.72         & 4.15        & 3.61        & 1.12        & 2.37         & 2.06         & 2.79        & 3.80 \\
$L_{1500}$\tablenotemark{a} & 6.1          & 5.5         & 0.41        & 4.6         & 0.39         & 0.24         & 3.68        & 0.21\\
$f_{\mathrm{Ly}\,{\alpha}}\tablenotemark{b}$  & 4.6$\pm$0.6  & 9.3$\pm$0.6 & 4.8$\pm$0.4 & 1.3$\pm$0.5 & 1.5$\pm$0.3  & 2.6$\pm$0.2  & 9.7$\pm$0.8 & 6.8$\pm$0.7\\
$L_{\mathrm{Ly}\,{\alpha}}\tablenotemark{c}$  & 2.1$\pm$0.3  & 1.6$\pm$0.1 & 1.4$\pm$0.1 & 2.9$\pm$1.1 & 1.4$\pm$0.28 & 3.4$\pm$0.3  & 29$\pm$2.4  & 6.9$\pm$0.7\\
$\mathit{SFR}_\mathrm{UV}\tablenotemark{d}$   & 6.4          & 5.8         & 0.44        & 4.8         & 0.41         & 0.25         & 3.9         & 0.22  \\
$\mathit{SFR}_\mathrm{Ly\,{\alpha}}\tablenotemark{d}$ & 0.19$\pm$0.03&0.15$\pm$0.01&0.13$\pm$0.01&0.26$\pm$0.09& 0.13$\pm$0.03& 0.31$\pm$0.03& 2.62$\pm$0.22& 0.35$\pm$0.04\\
$log_{10}(M_{*}/M_\odot)$ & $9.6 \pm 0.1$ & $8.7 \pm 0.2$ & $8.4 \pm 0.3$ & $9.7 \pm 0.1$ & $8.6 \pm 0.5$ & N/A & $9.4 \pm 0.3$ & $8.2 \pm 0.2$  \\
$R (\mathrm{kpc})$      & 2.8          & 2.2         & $<0.36$     & $<0.72$     & $<0.37$      & $<0.34$      & $<0.28$     & $<0.19$ \\ 
$\Sigma\tablenotemark{e}$          & 0.25         & 0.38        & $>1.08$      &  $>2.94$       &  $>0.95$ &  $>0.68$     &  $>15.8$    &  $>1.93$ \\
$M_{*}/\mathit{SFR}\tablenotemark{f}$  & 6.2   & 0.9        &  5.7        & 10.4         &  9.7         &  N/A         &  6.4        &  7.2  \\
\tableline
\end{tabular}

\tablenotetext{a}{$\times 10^{40}\,\mathrm{ergs\,s}^{-1}\,\mathrm{\AA}^{-1}$}

\tablenotetext{b}{$\times 10^{-17}\,\mathrm{ergs\,s^{-1}\,cm}^{-2}$}

\tablenotetext{c}{$\times 10^{41}\,\mathrm{ergs\,s}^{-1}$}

\tablenotetext{d}{$M_{\odot}\,\mathrm{yr}^{-1}$}

\tablenotetext{e}{$M_{\odot}\,\mathrm{yr}^{-1}\,\mathrm{kpc}^{-2}$}

\tablenotetext{f}{$\times 10^{8}\,\mathrm{yr}$}

\caption{\label{physics} Summary of physical properties derived for all LAEs at $z>1.5$ in our 
sample, and comparison with the mean values of the source at $z\sim 5.6$ found by \citet{Ellis} 
in Abell 2218 (rightmost column). From top to bottom: redshift, magnification factor $\mu$, 
unlensed UV luminosity ($L_\lambda$) at rest-frame $\lambda=$1500 \AA, observed integrated 
flux in the Lyman-$\alpha$ emission line, corresponding unlensed Lyman-$\alpha$ luminosity, 
star formation rate derived from the UV continuum ($\mathit{SFR}_\mathrm{UV}$) or the 
Lyman-$\alpha$ emission ($\mathit{SFR}_{\mathrm{Ly}_{\alpha}}$), stellar mass estimate 
from SED fitting (see Sect. \ref{smass}), intrinsic physical scale $R$ (see Sect. \ref{phys_scale}), 
star formation surface density $\Sigma$ and specific star formation rate.
}
\end{table*}

\subsection{Lyman-$\alpha$ emission surrounding C4}
\label{mapping}

\begin{figure*}
\mbox{\includegraphics[width=4cm]{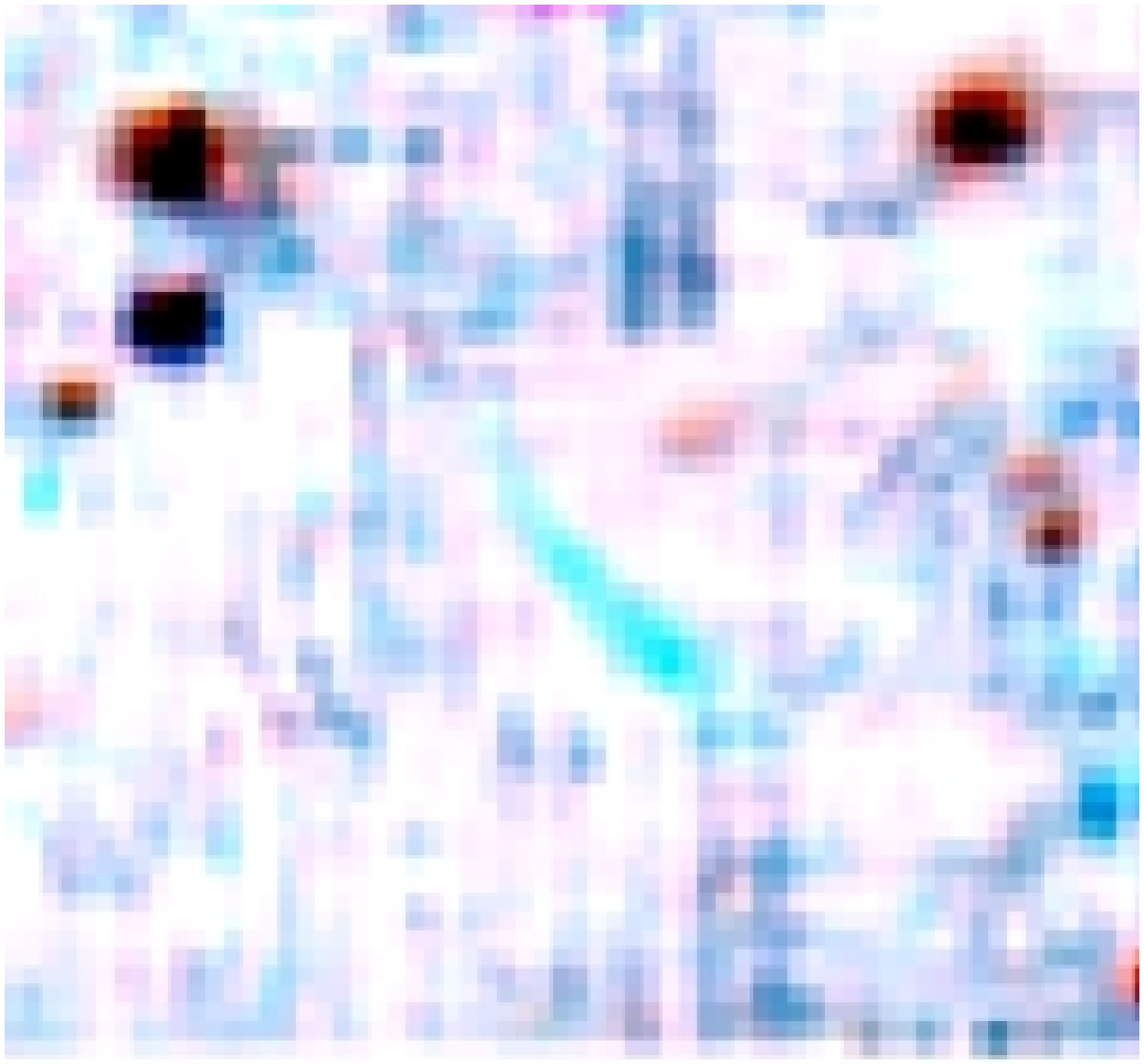}}
\hspace{.5cm}
\mbox{\includegraphics[width=8cm]{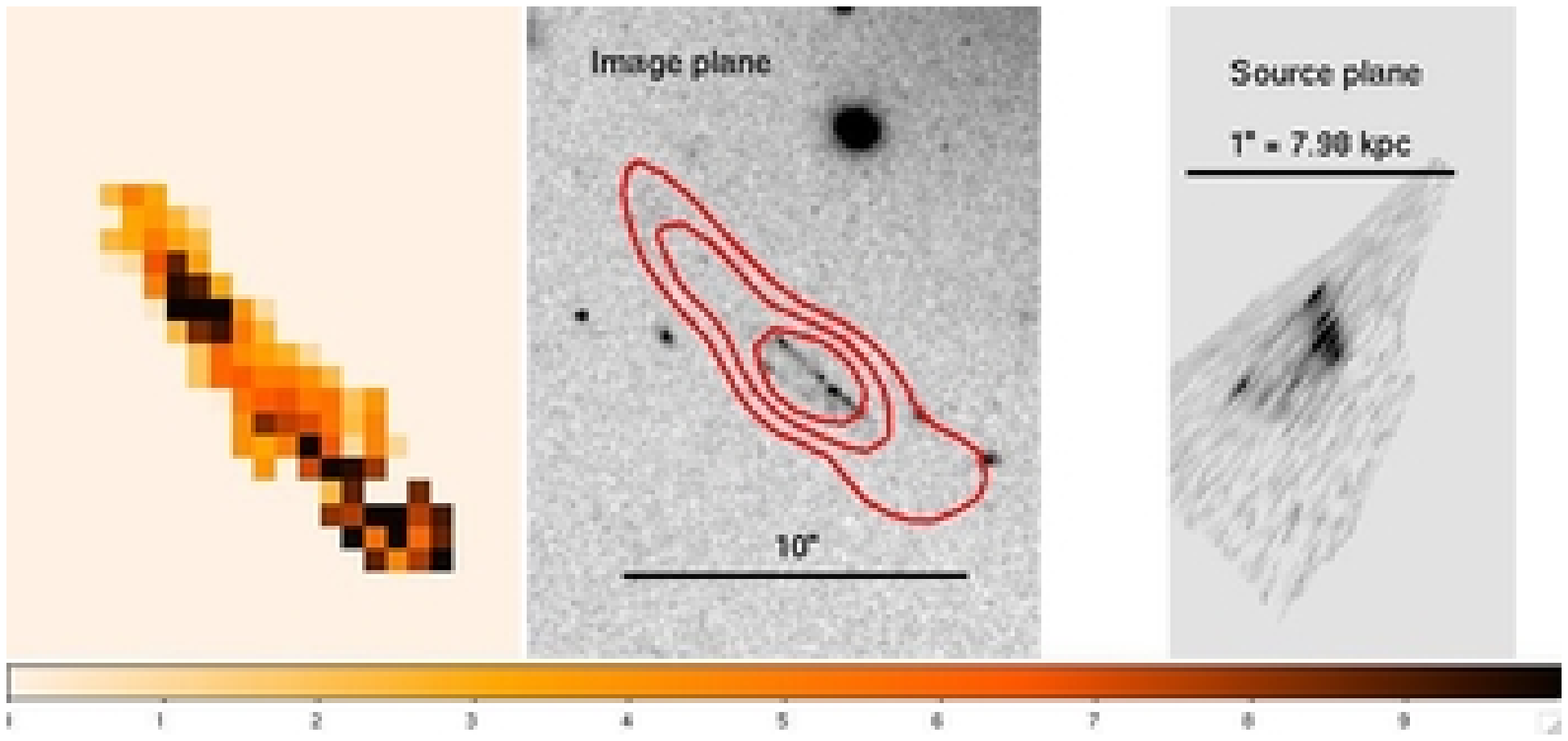}}
\caption{\label{C4ifu} (Left) 
Negative color image of a 40\arcsec side region around the source C4, produced by combining 
three different wavelength slices from the IFU datacube : a 15 \AA-wide region encompassing 
the Lyman-$\alpha$ emission (blue), and two broad wavelength regions at redder wavelengths 
(green and red). The extended emission around the arc C4 is clearly detected in the blue frame.}

(Right) Lyman-$\alpha$ emission properties of C4 observed with the IFU data. 
From left to right: rest-frame equivalent-width mapping of the Lyman-$\alpha$ emission (with 
corresponding color-scale), contours of Lyman-$\alpha$ emission overplotted on the HST image, 
$R$-band reconstruction of C4 in the source plane.
\end{figure*}

We now turn to a specific issue relating to the origin of the so-called Lyman-$\alpha$ blobs, of
which C4 at $z=2.63$ is the first lensed example. We have used the updated mass model 
to reconstruct the source-plane morphology of C4 . The high magnification ($\sim 4$ mags.)
enables us to resolve the source morphology at $\sim 30$ pc scales. The galaxy displays a 
bright component with an intrinsic elongated shape and several knots, at a scale length of 
$\sim$3 kpc  along the major axis. A fainter component having a similar shape and one bright 
knot is located $\sim 4.6$ kpc away (Fig. \ref{C4ifu}, fourth panel). Such an elongated shape 
and small physical scale is not uncommon among high redshift galaxies: \citet{Ravindranath} 
have measured that $\sim 50\%$ of galaxies in their sample of $z>2.5$ LBGs taken from 
GOODS show a bar-like morphology, and scale lengths of about $1.7-2.$ kpc. 

The strong Lyman-$\alpha$ emission of C4 detected with LRIS was also identified in the 
VIMOS/IFU data (Fig. \ref{C4ifu}, first panel). Compared with the observed size of the arc in the 
continuum image $R_{702}$ (4$\times$1.2\arcsec) this emission is significantly more 
extended ($\sim$15$\times$3\arcsec, third panel of Fig. \ref{C4ifu}). After correction for
an average magnification, this corresponds to an intrinsic scale of $\sim10$ kpc.

We use the $R_{702W}$ image to estimate the continuum flux of C4 at the wavelength of the 
Lyman-$\alpha$ emission, after applying a $k$-correction factor assuming a UV spectral slope of 2.0 
typical of starburst galaxies. By comparing the spatial coverage of the Lyman-$\alpha$ emission 
and the continuum, we construct a map of rest-frame Lyman-$\alpha$ equivalent width 
$W_0$ (Fig. \ref{C4ifu}, second and third panel). The $W_0$ values are typically in the range 
$5<W_0<10$ \AA\ in the region detected on the HST image. However, the map shows a 
significant increase of Lyman-$\alpha$ equivalent width, with values $W_0>10$ \AA, in the 
northern part of the arc, where the continuum is hardly visible in $R$ band. Because of the 
size of the blob and the presence of this stronger stellar continuum at the center of C4, the 
most probable mechanism producing such an extended Lyman-$\alpha$ emission is the 
presence of a superwind outflow originating from the central starburst \citet{Taniguchi, Wilman}.

\begin{figure*}
\centerline{\mbox{\includegraphics[width=7cm]{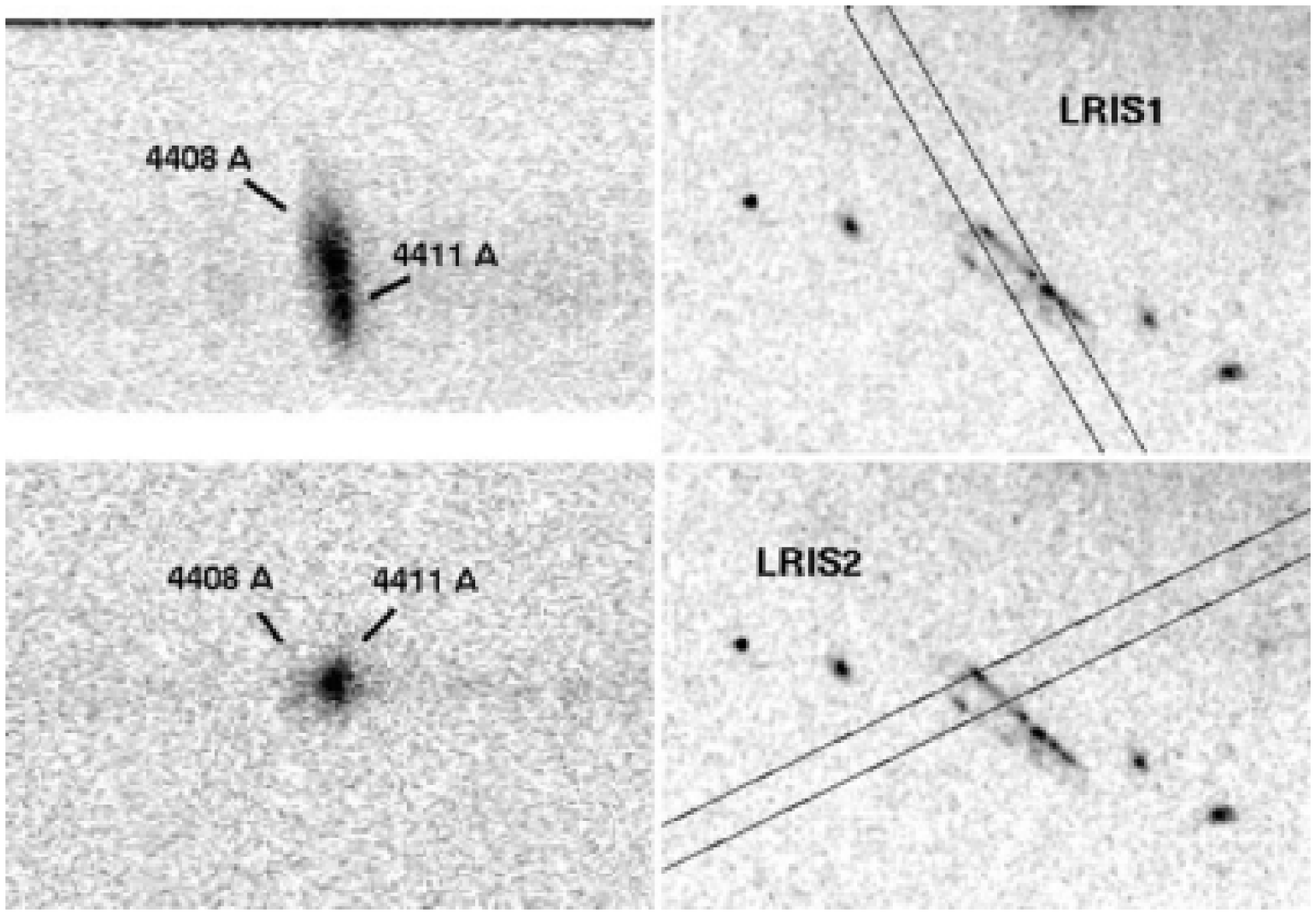}}}
\caption{\label{C4lris}
LRIS observations of C4. Left panels : closeups of each 2D spectrum showing the two components appearing in the Lyman-$\alpha$ emission. Right panels : corresponding LRIS slit configuration plotted over the HST image.
}
\end{figure*}

We also observed C4 with 2 different LRIS configurations, located across or along the longer 
dimension of the arc (Fig. \ref{C4lris}). In both cases, we observed two different components 
to the Lyman-$\alpha$ line on the 2D spectrum : a stronger emission centered around 
4411 \AA\ and a fainter emission region at a slightly bluer wavelength (4408 \AA\ ). The LRIS data
also shows a tilt in the emission region, due to an offset in the central peak of the Lyman-$\alpha$ 
line (top panel of Fig. \ref{C4lris}). These components may be related to the different star-forming 
regions identified morphologically in the HST image. Unfortunately, because of the poorer resolution, 
no similar offset in the central wavelength or variations in the line profile were detected in the IFU data.

The main difference between C4 and the very luminous Lyman-$\alpha$ blobs
is its very small physical size : 2.2 kpc, compared with the $R>30$ kpc selection criteria 
for Lyman-$\alpha$ blobs adopted by \citet{Matsuda}, along with the rest-frame equivalent width 
criterion $W_o>20$ \AA\ . For this reason, it is more likely that the associated physical processes 
 may be different, the size of C4 being more similar to extended Lyman-$\alpha$ emssions observed 
around star-forming galaxies. In this strongly lensed case, the effects of outflows on the Lyman-$\alpha$ 
emission can be studied in higher details by way of the lensing magnification, stretching the observed 
scales.

\subsection{Intrinsic properties of the $z=5.4$ source}

We finally turn to the most distant lensed source, C15, at $z=5.4$, and whether it is a further
example of the intriguing lensed source detected at $z$=5.6 by \citet{Ellis} in Abell 2218.

C15 has an intrinsic Lyman-$\alpha$ luminosity of $\sim 3\times 10^{42}\ \mathrm{ergs}\ \mathrm{s}^{-1}$. This is about 
5 times fainter than typical values found for $z\sim 5.7$ LAEs targetted by narrow-band searches 
\citep{Hu04,Shimasaku}, but not too dissimilar to those in the much fainter $z\sim6.5$ sample from 
\citet{Kashikawa} who used very deep ($\sim 10$ ks) spectroscopy with Keck and Subaru to confirm 
17 faint LAEs. 

All three images of this source are unresolved along the shear direction, and therefore suggest a very 
small physical scale $R\lesssim 300\,\mathrm{pc}$ in the source plane. More interestingly, we infer a
minimum star formation surface density of $\sim 16\,M_{\odot}\, \mathrm{kpc}^{-2}$, which is much 
higher than for any other object in our sample, making this source similar to the most active star 
forming regions at any redshift.

\begin{figure}
\centerline{\mbox{\plotone{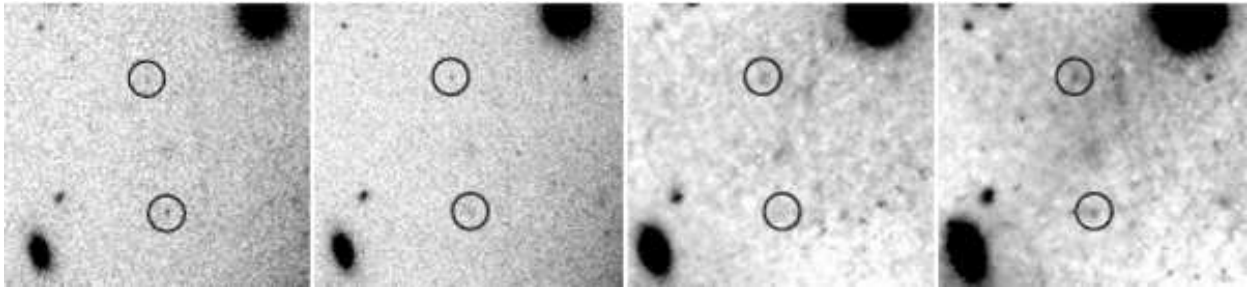}}}
\caption{\label{ellis01} HST broad-band detections of the $z\sim 5.6$ pair found by \citet{Ellis} in Abell 2218. From left to right:
$I_{F814W}$, $z_{850LP}$, $J_{F110W}$, $H_{F160W}$. North is up and East is left. The light from the nearby BCG 
(located towards the lower-right corner) has been subtracted from the NICMOS data for clarity.}
\end{figure}

The strong Lyman-$\alpha$ emission dominates the optical flux and in this respect it is very 
similar to the $z\sim5.6$ object found by \citet{Ellis} in the cluster Abell 2218. To compare these
sources, we have performed a re-analysis of the photometric data for that object, taking into
account HST data that arrived for Abell 2218 subsequent to the \citet{Ellis} analysis. This includes 
HST/ACS, NICMOS and Spitzer/IRAC observations presented by \citet{Kneib04} and \citet{Egami}. 
We find that the Ellis et al source is now detected in the $I_{F814W}$, $z_{F850LP}$, $J_{F110W}$ 
and $H_{160W}$ filters (Fig. \ref{ellis01} and Table \ref{ellis_table}). The source is faint (AB$\sim 26.6$)
and remains undetected with IRAC at 3.6 and 4.5 microns. However, the IRAC upper limits 
are quite high (around AB$\sim$24), due to the proximity of the BCG ($\sim 15\arcsec$) and coarse 
point spread function. 

In order to compare the revised data with that for C15, we compute and incorporate in the last column 
of Table \ref{physics} various physical properties for the Ellis et al. object alongside those for the 
other Lyman-$\alpha$ emitters. When deriving the stellar mass, we note the significant change
implied by the new data ($M_{*}\sim 10^{8}\,M_\odot$ c.f. $10^{6-7}\,M_\odot$ given by Ellis et al).
The major change arises from different assumptions : in order to reproduce the observed 
Lyman-$\alpha$ emission and the non-detection of the UV spectral continuum in LRIS, Ellis et al 
used the STARBURST99 spectrophotometric code \citet{S99} to derive an upper limit of 2 Myrs 
on the age of the source, assuming a constant $\mathit{SFR}$ and no extinction. 

Although the new HST photometry of this object is restricted to the rest-frame wavelength
$\lambda<2500$ \AA, the SED fitting method used in our reanalysis should be more reliable 
because it is based on fewer assumptions, especially on the star-formation history. Our best 
SED fitting model also predicts Spitzer/IRAC continuum fluxes consistent with the non-detection 
of this pair in the 3.6 $\mu m$ and 4.5 $\mu m$ bands.  

\begin{table}
\begin{tabular}{lllllll}
$V_{F606W}$ & $I_{F814W}$ & $z_{F850LP}$ & $J_{F110W}$ & $H_{F160W}$ & IRAC$_{3.6\mu m}$ & IRAC$_{4.5\mu m}$\\
\tableline
$>28.14$ & $26.94 \pm 0.18$ & $26.67 \pm 0.15$ &  $26.60 \pm 0.28 $  & $26.62 \pm 0.33 $ & $>24.1$ & $ >24.3$ \\
\end{tabular}
\caption{\label{ellis_table}Broad-band photometry of the $z=5.6$ galaxy discovered by \citet{Ellis}, 
measured on the new HST-ACS/NICMOS and Spitzer/IRAC images.}
\end{table}

Compared with the $z\sim 5.6$ galaxy from \citet{Ellis}, C15 is less magnified and is intrinsically more luminous and more massive, although with a similar physical size. The mass model predicts the 
source to be located $\sim 2.4$ kpc from the caustic line in the source plane. This suggests we are 
actually seeing a true small isolated object, since any similarly bright region close to C15 would also be 
highly magnified and multiply imaged.

Another difference between the two sources arises from the rest-frame UV stellar continuum, which is detected in the LRIS spectrum of the $z\sim 5.4$ source redward of the Lyman-$\alpha$ emission. 
Running the STARBURST99 code on this object in a similar way as done by \citet{Ellis}, 
the upper limit on the age is much larger, typically $50-100$ Myrs, with a constant 
$\textit{SFR}$ of 2.6 $M_\odot.\rm{yr}^{-1}$.  This gives a total mass of $\sim 1-3 \times 
10^8$ $M_\odot$, which is closer to the SED fitting estimate, in comparison with the source in 
Abell 2218.

At the time of its discovery, the \citet{Ellis} object revealed a very small and low-mass (10$^{6-7}$)
source at $z\sim 5.6$, which was also reported to be very young. With the new photometric 
reanalysis, and by finding another example of a LAE with similar size and Lyman-$\alpha$ flux, it 
is more probable that we are observing two 10$^8$ to 10$^9$ M$_\odot$ objects with very 
different star-formation 
histories. The Ellis et al. source previously formed the majority of its stellar mass and is having 
a very young burst, whereas C15 has been forming stars for a larger time scale.

\section{Discussion}

We have demonstrated in our survey of Abell 68 that, after applying magnification correction, 
we can reach LAEs with typical unlensed absolute AB magnitudes $\sim -20 < M_{\rm AB} < -16.5 $ 
in the rest-frame UV. This is about 2 magnitudes fainter than the UV selection criterion for typical
Lyman Break Galaxies used by \citet{Steidel03}, of $R_\mathrm{AB}<25.5$ at $z\sim 3$ and 
the faintest LAEs found with narrow-band searches \citep[e.g.,][]{Fynbo,Gawiser,Shimasaku}. 
Moreover, the lensed Lyman-$\alpha$ sources have much lower equivalent widths, similar to those 
seen in LBGs \citep{Shapley03}.

The derived properties of the Lyman-$\alpha$ emitters are dependent on the individual magnification factors $\mu$ 
used to correct each image for the lensing effects. The error estimates on $\mu$, computed by the bayesian optimization 
method, range from 1 to 16 \% and were quadratically added to the measurement errors in Table \ref{physics}, 
in the case of $L_{\mathrm{Ly}\,{\alpha}}$, $\mathit{SFR}_\mathrm{Ly\,{\alpha}}$ and $M_{*}$. In each case, the photometry 
is the dominant source of error. We acknowledge these values of $\mu$ are still dependent on the approach 
used to parametrize the mass model, for instance the use of the PIEMD profile for the dark matter haloes. Nevertheless, the Lyman-$\alpha$ emitters presented here     
are strongly lensed and have been identified in the central regions of the cluster, where the magnification factors 
are associated with the location of the critical lines. These critical lines are constrained by the same set of multiple 
images, quite independently of the parametrization used for the mass model (PIEMD or Navarro-Frenk-White (NFW) profile). 
Therefore, we are confident that the computed magnification factors are reliable.

Although such a lensing survey is not strictly flux-limited, we can use this unique probe to
gauge the properties of the faintest LAEs yet located at $z>2$. We find the typical stellar
masses are $\log_{10}(M_{*}/M_\odot)\sim 8.5$ to 9.5. These values are similar to the typical 
stellar masses of the bright LAEs from \citet{Gawiser}, found by selecting very high equivalent 
width ($W_0>150$) LAEs at z=3.1. Our LAEs have fainter UV and Lyman-$\alpha$ luminosities, 
and therefore a lower SFR, by typically 1 to 2 magnitudes. 
Even if the wavelength range covered by the broad-band photometry dominates 
the errors in the stellar masses derived in both surveys, this is indicative that, at comparable 
stellar masses, our LAEs are more quiescent than the objects found in narrow-band searches. 

\begin{figure*}
\centerline{\mbox{\includegraphics[width=10cm,angle=270]{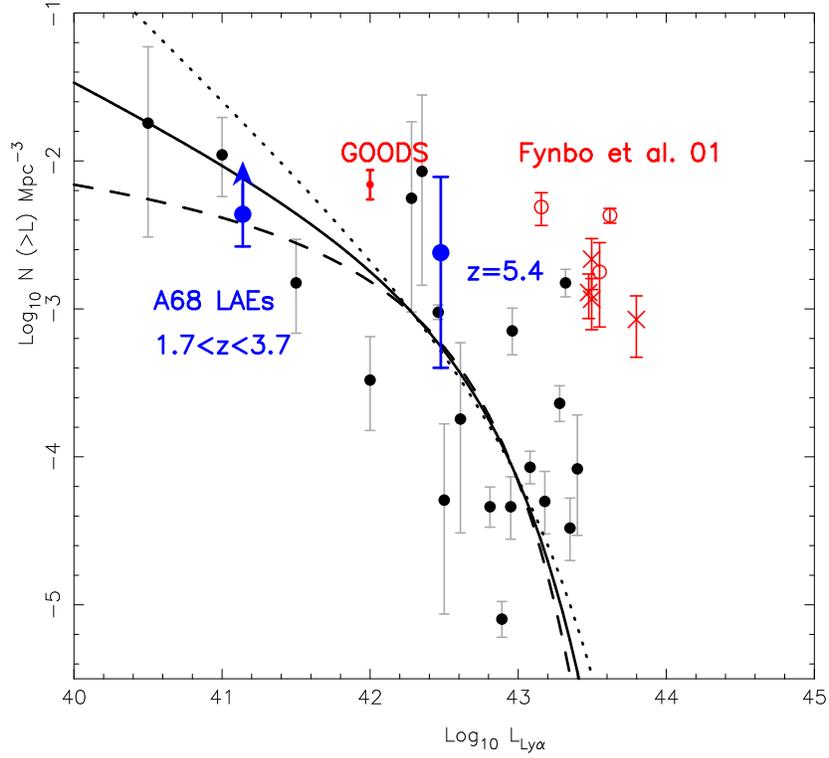}}}
\caption{\label{lf}
Cumulative luminosity function of Lyman-$\alpha$ emitters. Red datapoints are $z\sim3$ 
estimates from \citet{Fynbo01}, either in overdense regions (open circles) or blank fields (crosses). 
We also give recent results at $z\sim3.1$ from a Ly$\alpha$ survey in GOODS-South 
(Nilsson et al. 2007, to be submitted). Black points  correspond to the compilation of 
$z\sim 5$ surveys from \citet{Santos} and black curves are more recent fits to the 
$z\sim 5.7$ Luminosity Function by \citet{Shimasaku}. We overplot in blue number 
density estimates based on the sample of 6 Ly$\alpha$ emitters at $1.7<z<3.7$.
}
\end{figure*}

Recognizing the limitations of our sample, as a point of illustration, we compare on 
Fig. \ref{lf} the luminosity range of our lensed emitters with current constraints on the 
cumulative Lyman-$\alpha$ luminosity function at $z\sim3$, mostly based on narrow-band 
searches. The number of dedicated searches for LAEs at $z\sim 3$ from the literature is quite limited.
We report in this diagram the compilation of 5 samples, either in blank fields or overdense 
environments, presented by \citet{Fynbo01}, as well as an estimate of the number density 
from a Ly$\alpha$ survey in the GOODS-south field (\citet{Nilsson} and Nilsson et al. 2007, to be submitted).
Based on our subsample of 6 emitters at $2\lesssim z \lesssim 3.7$ and the surface area probed 
 in the high magnification region ($\mu>0.75$ mag.) of the source plane, we estimate a number density 
of 4.3$^{+2.6}_{-1.7}\times 10^{-3}$ Mpc$^{-3}$ down to $L=1.4 \times 10^{41}\ \rm{ergs}\ \rm{s}^{-1}$.  
Because our spectroscopy only sparsely covered this region, we report this value as a lower 
limit in Fig. \ref{lf}. Although we acknowledge the limits due to very small statistics and the 
small areas probed in these lensed surveys, our sample of $z\sim 3$ emitters gives constraints 
at fainter luminosities that any of the narrow-band searches at this redshift. More observations 
are however needed in order to better constrain the faint-end slope of this luminosity function.

An interesting question is the likelihood of finding highly magnified sources at $z\sim5.5$ like C15 and
the source reported by \citet{Ellis} : we can estimate this through the space density of 
$R$-dropouts in our magnified field of view. Based on our photometric $RIz$ color-color selection 
technique (see Sect. \ref{riz}), we identified one source in the field of view covered 
by the HST image. Since it is detected at $\sim 8\,\sigma$ level in the $I$-band filter, where 
$R$-dropouts are brighter, and has a magnification factor of $\mu \sim 2.5$, we computed the 
effective covolume in the source plane having $\mu>0.75$, taking into account the reduction of 
the surface area due to lensing effects. We derive an effective covolume of 420 Mpc$^3$, in the 
redshift range $5<z<6$ probed by our set of filters. Assuming Poisson noise statistics, we obtain 
a rough estimate of $2.4^{+5.4}_{-2.0}$ $\times10^{-3}\,\mathrm{Mpc}^{-3}$ for the number 
counts of $R$-dropouts. Although cosmic variance effects can be quite large  when probing 
such a small volume, this result is in good agreement with the space density 
of low-luminosity Lyman-$\alpha$ sources at $z\sim5$, on the above-mentioned diagram (Fig. \ref{lf}).

From these simple calculations, C15 does not seem to be a serendipitous case of low-luminosity $z\sim 5$ 
LAE, since we would expect to find one such source in the magnified region of the cluster. By observing 
a larger number of lensed fields, such as the new HST clusters imaged by ACS, we could build a more 
significant sample of similar objects and compare their physical properties.

\section{Conclusions}

We have performed a spectroscopic analysis of background sources in the field of the 
massive cluster Abell 68 and identified 26 lensed images in the range $0.3<z<5.5$, including 
7 Lyman-$\alpha$ emitters at $z>1.7$. Using the new redshift measurements 
and identification of new multiple-image systems, we perform a precise modeling of the cluster mass distribution with 5 spectroscopic systems, and also predict the redshift and counter-images for 
two remaining systems. This makes Abell 68 one of the best-modelled lensing clusters, along 
with Abell 1689 and Abell 2218, allowing for the precise measurement of its dark-matter distribution.

We derived the star formation rates, stellar masses and physical scales for our sample of 
high-redshift Lyman-$\alpha$  emitters. The broad-band luminosities of these objects is comparable 
to the faint LAEs found in deep narrow-band searches, but their equivalent widths are much lower, 
making them 1 to 2 magnitudes fainter in Lyman-$\alpha$ luminosity. Two of these sources show 
a more extended Lyman-$\alpha$ emission region, when compared to 
the stellar continuum in the HST image. For one of them, we demonstrated how we use IFU data 
to probe regions with distinct Lyman-$\alpha$ equivalent widths. The stretch provided by lensing
enables us to characterize small regions of Lyman-$\alpha$ emission which would otherwise be 
beyond reach of ground based Integral Field Instruments. Although the large equivalent widths 
are comparable to giant Lyman-$\alpha$ blobs observed around massive forming galaxies, we 
interpret the small physical scales of these lensed emissions as outflows originating from 
a central starburst.

The highest redshift $z\sim 5.4$ multiple-image source of this sample is very similar to the 
pair of strongly-lensed images identified by \citet{Ellis} in Abell 2218, in terms of magnification 
and physical size, albeit being intrinsically more massive and more luminous. We therefore 
expect to detect it in IRAC images of similar depth as the data presented in \citet{Egami}, 
in the first two channels of this instrument. Such measurements would tighten the constraints 
on the stellar mass of this source, by reducing the degeneracies in the other model parameters, and 
also provide additional information on its age and star-formation history, when combined with 
the properties of the Lyman-$\alpha$ emission.

Our survey provides the first tentative indications of the density of faint Lyman-$\alpha$ 
emitters at $z\sim3$, down to unlensed fluxes of $\sim 2\times10^{41}$. Although our
survey is not flux-limited nor complete in any formal sense, the cumulative Lyman-$\alpha$ 
luminosity function we derive indicates the promise of a dedicated search for lensed 
Lyman-$\alpha$ emitters through larger samples of well-mapped clusters now being
surveyed with HST, and the ability of this approach to complement narrow-band searches 
carried out in blank fields. 

\acknowledgments

We thank the anonymous referee for his helpful comments, and acknowledge helpful discussions with Johan Fynbo, Kim Nilsson, Fabrice Lamareille, Mark Swinbank and Tommaso Treu. JR is grateful to Caltech for financial support. The Dark Cosmology Centre is funded by the Danish National Research Foundation.

\end{document}